\documentclass{emulateapj}
\usepackage[figuresright]{rotating}

\def\arcs{$''$}

\def\hub{\ifmmode H_\circ\else H$_\circ$\fi}

\shorttitle{Star clusters in M33} \shortauthors{Ma}

\begin{document}

\slugcomment{AJ, in press}
\title{NEW $UBVRI$ PHOTOMETRY of 234 M33 STAR CLUSTERS}

\author{Jun Ma\altaffilmark{1,2}}

\altaffiltext{1}{National Astronomical Observatories, Chinese Academy of Sciences, A20 Datun Road, Chaoyang District, Beijing 100012, China; majun@nao.cas.cn}

\altaffiltext{2}{Key Laboratory of Optical Astronomy, National Astronomical Observatories, Chinese Academy of Sciences, Beijing 100012, China}

\begin{abstract}
This is the second paper of our series. In this paper, we present $UBVRI$ photometry for 234 star clusters in the field of M33. For most of these star clusters, there is photometry in only two bands in previous studies. The photometry of these star clusters is performed using archival images from the Local Group Galaxies Survey, which covers 0.8 deg$^2$ along the major axis of M33. Detailed comparisons show that, in general, our photometry is consistent with previous measurements, especially, our photometry is in good agreement with Zloczewski \& Kaluzny. Combined with the star clusters' photometry in previous studies, we present some results: none of the M33 youngest clusters ($\sim 10^7$~yr) have masses approaching $10^5$~$M_{\odot}$; comparisons with models of simple stellar populations suggest a large range of ages of M33 star clusters, and some as old as the Galactic globular clusters.
\end{abstract}


\section{INTRODUCTION}
\label{s:intro}

Star clusters are an important tool for studying the star formation histories of galaxies. They represent, in distinct and luminous ``packets,'' single-age and single-abundance points and encapsulate at least a partial history of the parent galaxy's evolution.

M33 is a small Scd Local Group galaxy. It is located $\sim 809 \pm 24$ kpc from us (distance modulus $(m-M)_0=24.54\pm0.06$; McConnachie et al. 2004, 2005). M33 is interesting and important because it represents an intermediate morphological type between the largest ``early-type'' spirals and the dwarf irregulars in the Local Group. So, it can provide an important link between the star cluster populations of earlier-type spirals (Milky Way and M31) and the numerous nearby later-type dwarf galaxies.

In the pioneering work of M33 star clusters, \citet{hiltner60} presented photometry for 23 M33 star cluster candidates and 23 M31 globular clusters in the $UBV$ passbands using photographic plates taken with the Mt. Wilson 100-inch (2.5-m) telescope. And he found that, except for five of them, the star clusters in M33 are bluer and fainter than those in M31. At the same time, \citet{KM60} identified four M33 star clusters for which they gave $PV$ photometry. Then, \citet{MD78}
detected 58 star cluster candidates in M33 based on a baked IIIa-J+GG385 plate covering a field of about $1^\circ$ in diameter, including $B$ photometry of them. The most comprehensive catalog of nonstellar objects in M33 was compiled by \citet{CS82,CS88}, who detected 250 nonstellar objects by visually examining a single photographic plate taken at the Ritchey-Chrestien focus of the 4-m telescope at Kitt Peak National Observatory. These authors obtained ground-based $BVI$ photometry of 106 of these objects, which they believe to be star clusters. However, the star cluster candidates detected by these authors are limited to the outer part of M33.

The first survey for M33 star clusters based on CCD imaging was performed by \citet{Mochejska98}, using the data collected in the DIRECT project \citep{Kaluzny98,Stanek98}. These authors detected 51 globular cluster candidates in M33, 32 of which were not previously cataloged. These globular cluster candidates covered the central region of M33. In addition, \citet{Mochejska98} presented $BVI$ photometry for these globular cluster candidates.

Since the pioneering work of \citet{CBF99a}, the era of detecting and studying M33 star clusters based on the images taken with {\sl Hubble Space Telescope} ({\sl HST}) has begun \citep{CBF99a,CBF99b,CBF99c,CBF01,CBF02,Bedin05,PL07,Sarajedini07, Sarajedini98, Sarajedini00, Stonkute08,PP09,Huxor09,Roman09,ZK09}. The {\sl HST} resolution makes it easy to distinguish individual stars from star clusters at the distance of M33. So, M33 star clusters identified with {\sl HST} images are much less likely to be contaminated by other extended sources, such as a background galaxy or an HII region \citep[see][for details]{PL07}.

\citet{Ma01,Ma02a,Ma02b,Ma02c,Ma02d,Ma04a,Ma04b} constructed spectral energy distributions in 13 intermediate filters of the Beijing--Arizona--Taiwan--Connecticut photometric system for known M33 star clusters and star cluster candidates, and estimated star cluster properties.

In order to construct a single master catalog incorporating the entries in all of the individual catalogs including all known properties of each star cluster, \citet{sara07} merged all of the above-mentioned catalogs before 2007, for a summary of the properties of all of these catalogs. This catalog contains 451 star cluster candidates, of which 255 are confirmed star clusters based on the {\sl HST} and high-resolution ground-based imaging. The positions of the star clusters in \citet{sara07} were transformed to the J2000.0 epoch and refined using the Local Group Galaxies Survey (LGGS; Massey et al. 2006).

Very recently, some authors used the images observed with the MegaCam camera on the 3.6-m Canada--France--Hawaii Telescope (CFHT/MegaCam) to detect star clusters in M33 \citep{ZK08,Roman10}. \citet{Sharina10} presented the evolutionary parameters of 15 GCs in M33 based on the results of medium-resolution spectroscopy obtained at the Special Astrophysical Observatory 6-m telescope. Most recently, \citet{Robert11} searched for outer halo star clusters in M33 based on CFHT/MegaCam imaging as part of the Pan-Andromeda Archaeological Survey.

\citet{Ma12} (Paper I) presented $UBVRI$ photometry of 392 objects (277 star clusters and 115 star cluster candidates) in the field of M33, using the images of the LGGS (Massey et al. 2006). And he also provided properties of M33 star clusters such as their color--magnitude diagram and color--color diagram.

In this paper, we perform aperture photometry of 234 M33 star clusters based on the M33 images of the LGGS. These sample star clusters are selected from \citet{PL07}, \citet{Roman09} and \citet{ZK09}. This paper is organized as follows. Section 2 describes the sample star cluster selection and $UBVRI$ photometry. In Section 3, we present an analysis of the star cluster properties. Lastly, our conclusions are presented in Section 4.

\section{DATA}
\label{s:data}

\subsection{Sample}
\label{s:samp}

In Paper I, we presented an updated $UBVRI$ photometric catalog containing 392 star clusters and star cluster candidates in the field of M33 which were selected from the most recent star cluster catalog of \citet{sara07}. And we also provided properties of M33 star clusters such as their color--magnitude diagram (CMD) and color--color diagram combined with the photometry of M33 star clusters from \citet{PL07}, \citet{Roman09} and \citet{ZK09}. However, we found that most of M33 star clusters from \citet{Roman09} and \citet{ZK09} have photometry in only two bands $V$ and $I$. In the color--color diagram of Paper I, there are only $\sim 300$ M33 star clusters, since $\sim 200$ star clusters have no $B-V$ data. So, integrated magnitudes of these star clusters in $B$ and $V$ bands are emergently needed for studying the properties of M33 star clusters. In this paper, we will provide $UBVRI$ photometry of M33 star clusters from  \citet{PL07}, \citet{Roman09} and \citet{ZK09}. \citet{PL07} found 104 star clusters in the {\sl HST}/WFPC 2 archive images of 24 fields that were not included in previous studies, of which 32 star clusters are newly detected. \citet{ZK08} presented a catalog of 4780 extended sources in a 1 ${\rm deg^2}$ region around M33 including 3554 new star cluster candidates using the MegaCam camera on the CFHT. \citet{ZK09} used deep Advanced Camera for Surveys Wide Field Channel (ACS/WFC) images of M33 to check the nature of extended objects detected by \citet{ZK08}, and found that 24 star cluster candidates were confirmed to genuine compact star clusters. In addition, \citet{ZK09} detected 91 new star clusters based on these deep ASC/WFC images of M33, and provided integrated magnitudes and angular sizes for all these 115 star clusters. \citet{Roman09} presented integrated photometry and color--magnitude diagrams for 161 star clusters in M33 based on the ACS/WFC images, of which 115 were previously uncataloged. By cross-checking with the updated photometric catalog of M33 star cluster and candidate in Paper I, we found that, the photometry of 36 star clusters of \citet{PL07} was not presented in Paper I, of which the 32 star clusters were newly detected by \citet{PL07} and the remaining four were detected by previous studies. The three of the four star clusters were included in \citet{sara07} and were classified as `Stellar' (objects 69, 293 and 279 of Sarajedini \& Mancone 2007 which being called star clusters 36, 195 and 197 in Park \& Lee 2007, respectively), and the remaining one is star cluster 75 in \citet{PL07}. The photometry of 118 star clusters of \citet{Roman09} was not presented in Paper I, of which 115 star clusters were newly detected by \citet{Roman09} based on the ACS/WFC images, and the remaining three star clusters were included in \citet{sara07} which were classified as `Galaxy' or `Stellar' (objects 57, 62 and 69 of Sarajedini \& Mancone 2007 which being called star clusters 27, 34 and 38 in San Roman et al. 2009). The photometry of all star clusters of \citet{ZK09} was not presented in Paper I, of which one star cluster was included in \citet{sara07} and was classified as `Galaxy' (object 57 of Sarajedini \& Mancone 2007 which being called 33-3-021 in Zloczewski \& Kaluzny 2009). So, in this paper, we will perform photometry for the M33 star clusters in \citet{PL07}, \citet{ZK09} and \citet{Roman09} that were not presented in Paper I. Altogether, there are 269 star clusters combining the star clusters from \citet{PL07}, \citet{ZK09} and \citet{Roman09}. However, by cross-checking the coordinates of the star clusters of \citet{PL07}, \citet{ZK09} and \citet{Roman09}, and by checking the images of star clusters from the LGGS images, we found that, star clusters 7, 10, 14, and 18 of \citet{PL07} are the same objects with star clusters 33, 51, 59, and 64 of \citet{Roman09}, respectively. In addition, there are 18 common star clusters between \citet{ZK09} and \citet{Roman09} \citep[see Table 3 of][]{Roman09}. When we do photometry of the sample star clusters in this paper, we found that, there is nothing in the position of star cluster 195 of \citet{PL07} (i.e., no. 17 of Bedin et al. 2005), which was named object 293 in \citet{sara07} and was classified as `Stellar' by \citet{sara07}. We also found that, in the LGGS images of M33, (1) there are some bright objects near star cluster 12 of \citet{PL07}; (2) there is a bright object near star clusters 23 and 32 of \citet{PL07}, respectively; (3) there is a bright object very near star clusters 15, 114 and 141 of \citet{Roman09}, respectively; (4) there are three bright objects near star cluster 143 of \citet{Roman09}; (5) there is a very close object to star clusters ZK-21, ZK-22, ZK-28, ZK-66 and ZK-72 of \citet{ZK09}, respectively. The photometry of these 13 star clusters cannot be determined accurately in this paper. So, this paper will present homogeneous $UBVRI$ photometries for 234 star clusters in M33 using the images of the LGGS (see details about the LGGS in Paper I).

\begin{figure*}
\centerline{\includegraphics[scale=0.8,angle=-90]{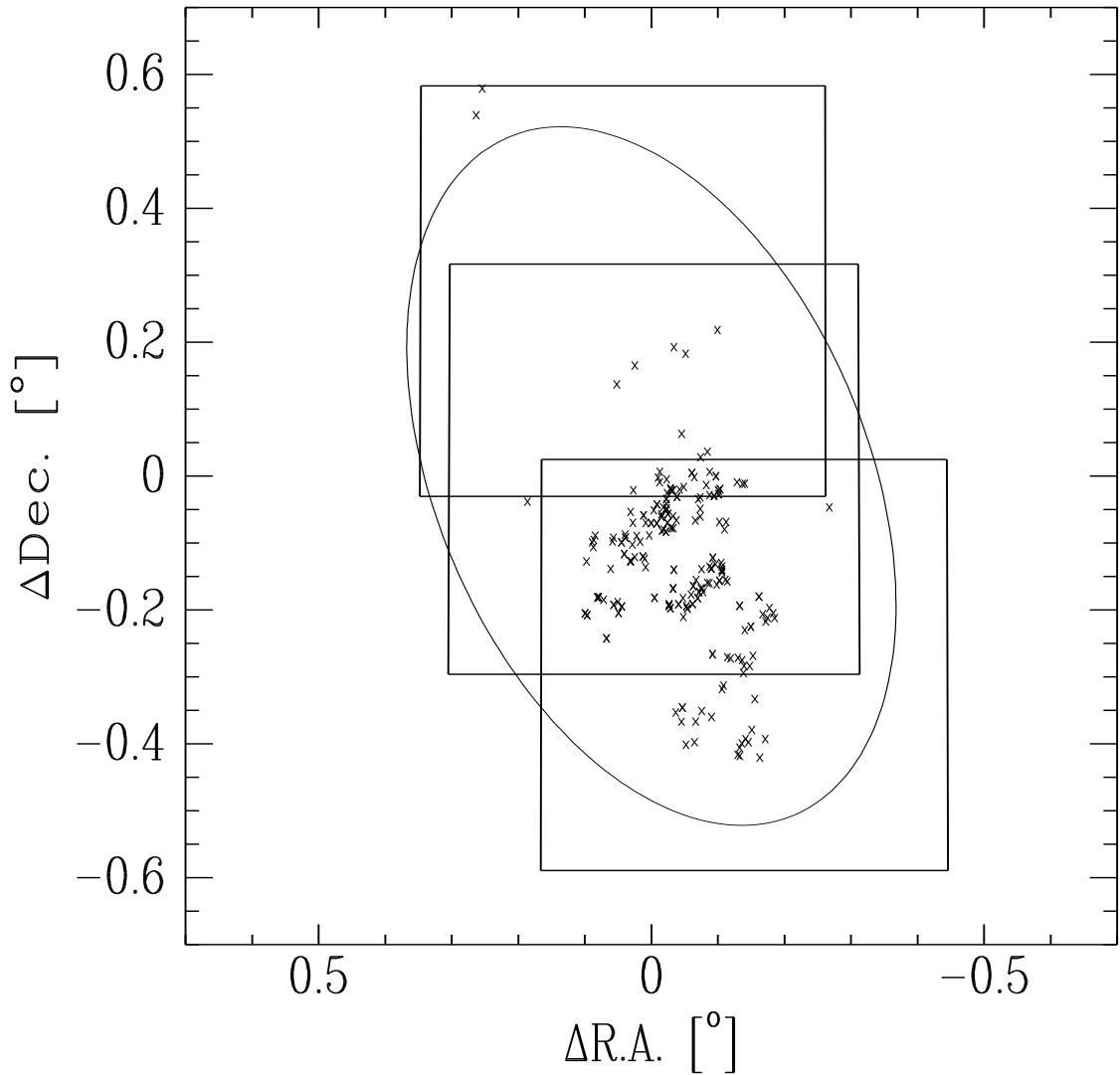}}
\caption[]{Spatial distribution of the 234 star clusters of M33 which were selected from \citet{PL07}, \citet{Roman09}, and \citet{ZK09} and their loci in the LGGS fields. We determined the photometry for these star clusters based on the LGGS archival images of M33 in the $UBVRI$ bands. The large ellipse is the $D_{25}$ boundary of the M33 disk \citep{Vaucouleurs91}. The three large squares are the LGGS fields.}
%\label{fig1}
\end{figure*}

\subsection{Photometry}
\label{s:phot}

We used the LGGS archival images of M33 in the $UBVRI$ bands to do photometry (see details in Paper I). We performed aperture photometry of the 234 M33 star clusters found in the LGGS images in all of the $UBVRI$ bands to provide a comprehensive and homogeneous photometric catalog for them. The photometry routine we used is {\sc iraf/daophot} \citep{stet}. The photometric process used in this paper is the same as in Paper I. We have checked the aperture of every sample star cluster considered here by visual examination to make sure that it was not too large (to avoid contamination from other sources). The aperture photometry of star clusters was transformed to the standard system using transformation (constant offsets neglecting color term) derived based on aperture photometry of stars whose $UBVRI$ magnitudes were published by \citet{massey}, who calibrated their photometry with standard stars of \citet{lan92}. Finally, except for star cluster 27 of \citet{Roman09} (i.e., SR27, which was named 33-3-021 in Zloczewski \& Kaluzny 2009) and ZK-82 of \citet{ZK09} in the $I$ band, and ZK-90 of \citet{ZK09} in the $U$ and $I$ bands, we obtained photometry for 234 star clusters in the individual $UBVRI$ bands. SR27 falls in the gap of the LGGS image in the $I$ band, and ZK-82 and ZK-90 in the $I$ band fall in the bleeding CCD column of a saturated star, and ZK-90 in the $U$ band does not lie in the LGGS image. Table 1 lists our new $UBVRI$ magnitudes and the aperture radii used (we adopted 0.258\arcs pixel$^{-1}$ from the image header), with errors given by {\sc iraf/daophot}. The star cluster names follow the naming convention of \citet{sara07} (i.e., ${\rm SM\times\times\times}$), \citet{PL07} (i.e., ${\rm PL\times\times\times}$), \citet{Roman09} (i.e., ${\rm SR\times\times\times}$), and \citet{ZK09}. In addition, we also list the reddening values of the sample star clusters in Table 1 (see Section 3.1 for details). In Table 1, $R_C$ and $I_C$ mean that $RI$ magnitudes are on Johnson--Kron--Cousins system.

To examine the quality and reliability of our photometry, we compared the aperture magnitudes of the 234 star clusters obtained here with previous photometry of \citet{PL07}, \citet{Roman09}, and \citet{ZK09}. There are eight star clusters, of which the magnitude scatters in the $V$ band between this study and previous studies of \citet{PL07} and \citet{Roman09} are larger than 1.0 mag, i.e., our magnitudes are fainter than those obtained by \citet{PL07} and \citet{Roman09}. We listed the comparison between this study and previous studies of $V$ photometry for these eight star clusters in Table 2. We also plotted their images in Figure 2, in which the circles are photometric apertures adopted here. From this figure, we can see that nearly all these star clusters are close to one or more bright sources. If photometric apertures are larger than the values adopted here, the light from these bright sources will not be excluded. As we know that, in \citet{PL07}, the $BVI$ integrated aperture photometry of M33 star clusters, which is included in $50'\times80'$ field of M33 based on CCD images taken with the CFH12k mosaic camera at the CFHT, is derived with an aperture of $r=4.0''$ for $V$ magnitude measurement and an aperture of $r=2.0''$ for the measurement of color. \citet{Roman09} derived integrated photometry and color-magnitude diagrams (CMDs) for 161 star clusters in M33 using the ACS/WFC images. These authors adopted an aperture radius of $r=2.2''$ for $V$ magnitude measurements and $r=1.5''$ for the colors. For these eight star clusters, a large scatter in the $V$ photometric measurement between this study and previous studies \citep{PL07,Roman09} mainly results from different photometric aperture sizes adopted by different authors (see Paper I for details). Figures 3--5 show the comparison of our photometry of the star clusters with previous photometry of \citet{PL07}, \citet{Roman09} and \citet{ZK09}. PL197 is not included in the figure of $\Delta V$ comparison of Figure 3 because of too large value of $\Delta V$ to be drawn in the figure. In addition, in Figure 5 (and Figures 6, 8 and 9 below), we have transformed the ACS/WFC magnitudes in F475W, F606W and F814W bands to the Johnson-Cousins $B$, $V$ and $I$ magnitudes using the color-dependent synthetic transformations given by Sirianni et al. (2005).

\begin{figure*}
\centerline{
\includegraphics[scale=0.8,angle=0]{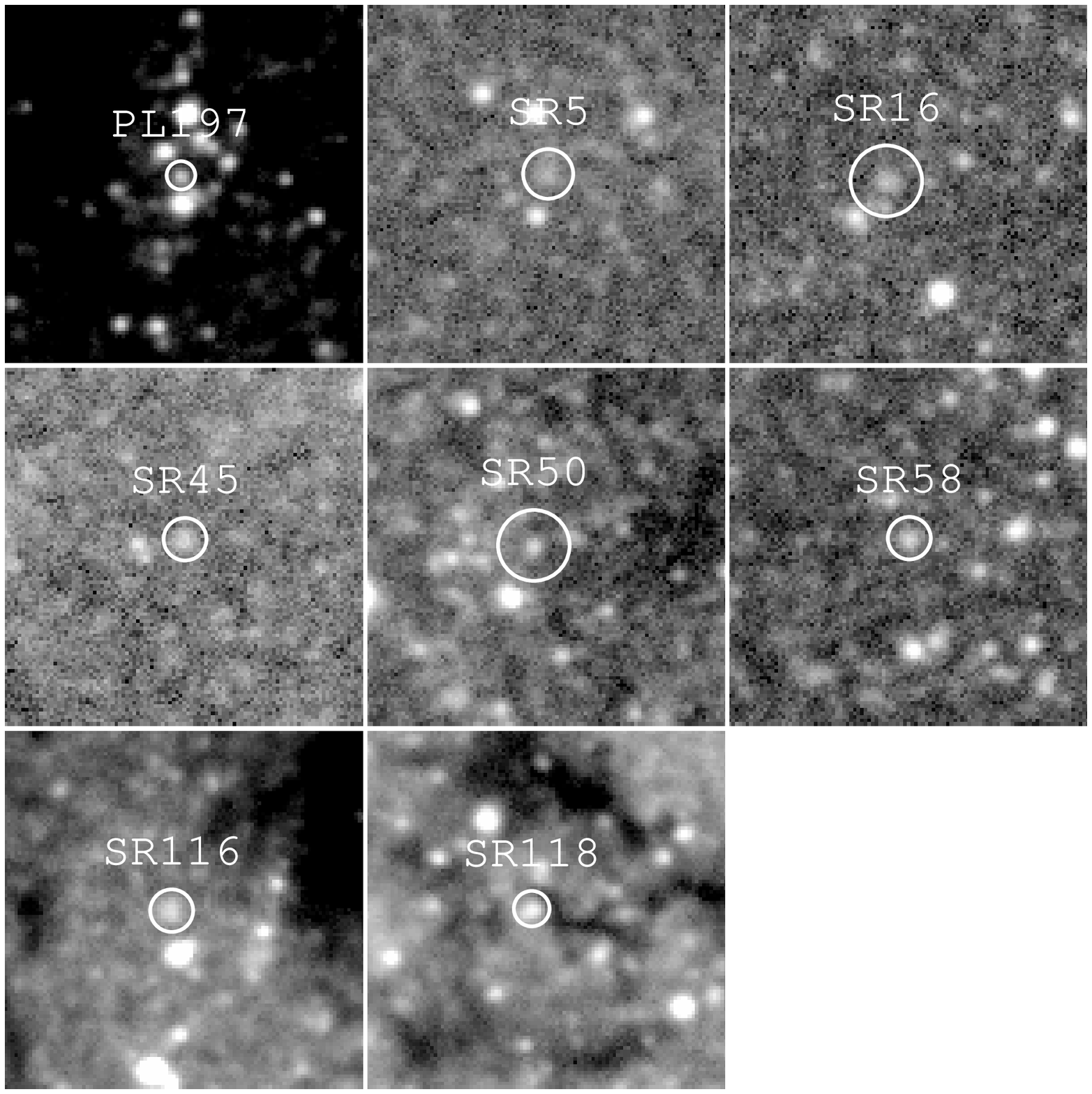}}
\vspace{-6cm}
\caption{Finding charts of eight star clusters in the LGGS $V$ band, of which the magnitude scatters in the $V$ band between this and those studies of \citet{PL07} and \citet{Roman09} are larger than 1.0 mag, i.e., our measurements are fainter than those in \citet{PL07} and \citet{Roman09}. The circles are photometric apertures adopted in this paper.}
%\label{fig2}
\end{figure*}

\begin{figure*}
\centerline{
\includegraphics[height=140mm,angle=-90]{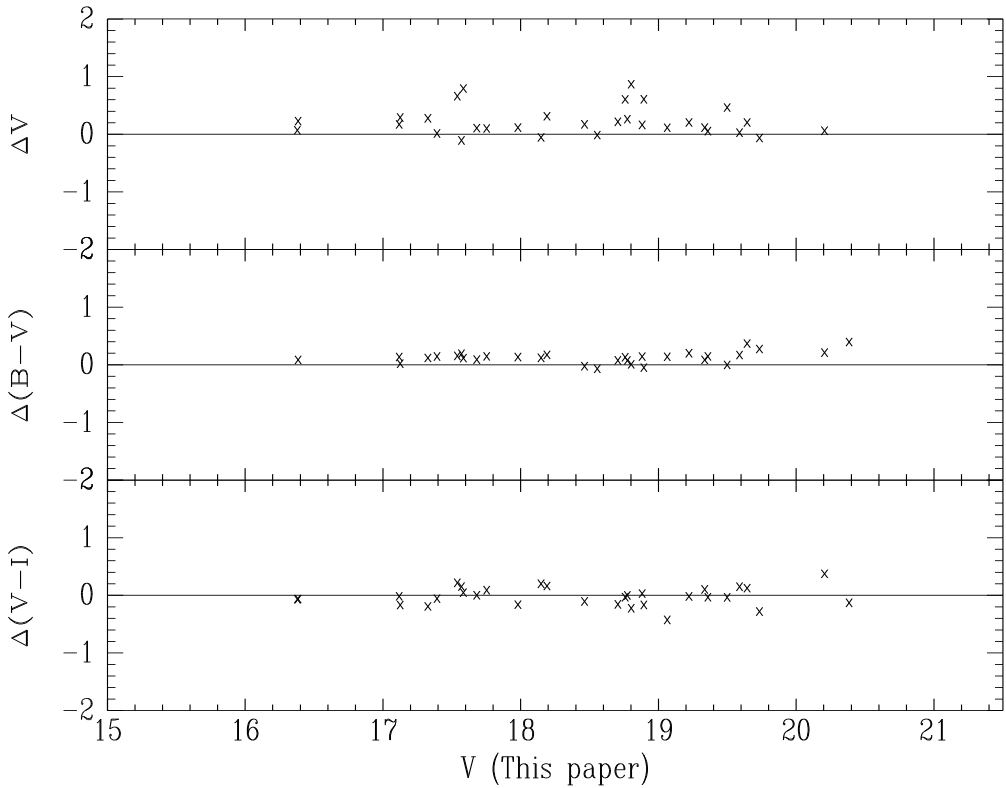}}
\vspace{0.5cm}
\caption{Comparisons of our photometry of M33 star clusters in the $UBVRI$ bands with previous photometry in \citet{PL07}. The photometric offsets and rms scatter of the differences between their measurements and our new magnitudes are: $\Delta V = 0.306\pm0.089$ with $\sigma=0.498$, $\Delta (B-V) = 0.126\pm0.019$ with $\sigma=0.102$, and $\Delta (V-I) = -0.023\pm0.030$ with $\sigma=0.162$ (this study minus Park \& Lee 2007).}
\end{figure*}

\begin{figure*}
\centerline{
\includegraphics[height=140mm,angle=-90]{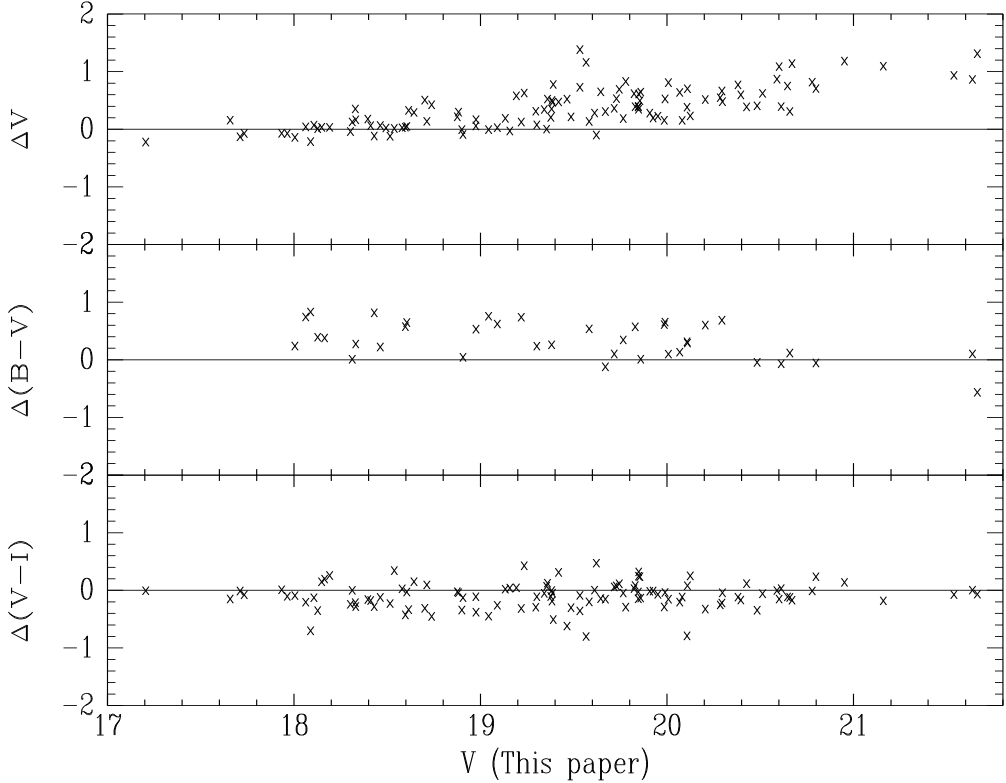}}
\vspace{0.5cm}
\caption{Comparisons of our photometry of M33 star clusters in the $UBVRI$ bands with previous photometry in \citet{Roman09}. The photometric offsets and rms scatter of the differences between their measurements and our new magnitudes are: $\Delta V = 0.363\pm0.033$ with $\sigma=0.352$, $\Delta (B-V) = 0.332\pm0.052$ with $\sigma=0.316$, and $\Delta (V-I) = -0.100\pm0.021$ with $\sigma=0.244$ (this study minus San Roman et al. 2009).}
\end{figure*}

\begin{figure*}
\centerline{
\includegraphics[height=140mm,angle=-90]{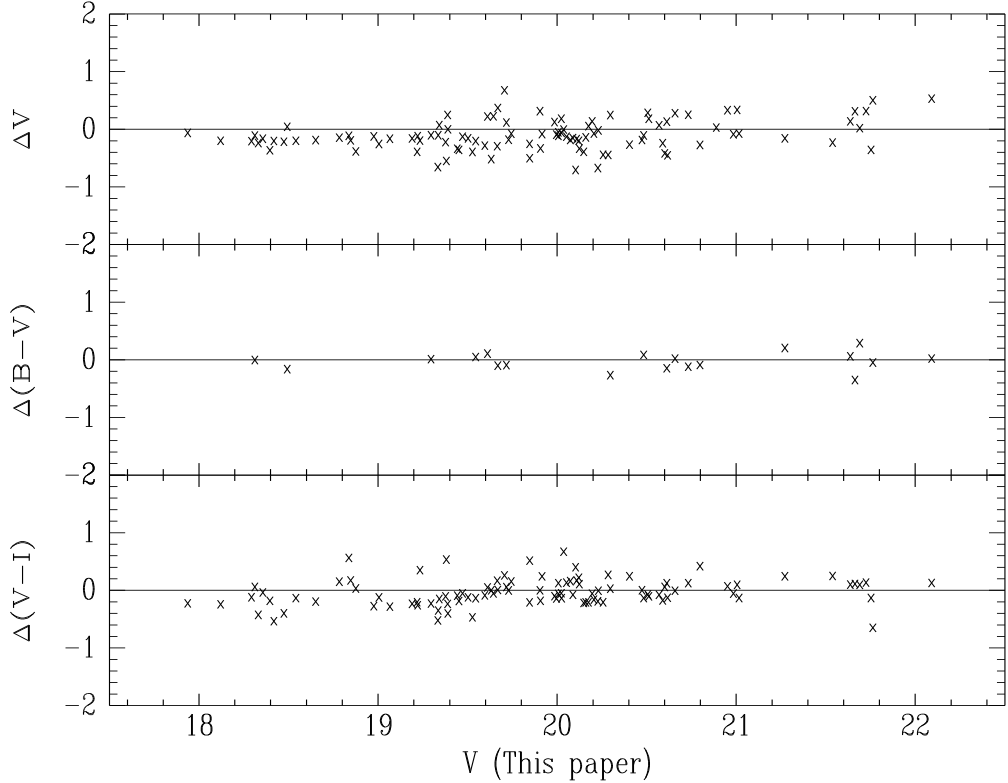}}
\vspace{0.5cm}
\caption{Comparisons of our photometry of M33 star clusters in the $UBVRI$ bands with previous photometry in \citet{ZK09}. The photometric offsets and rms scatter of the differences between their measurements and our new magnitudes are: $\Delta V = -0.103\pm0.026$ with $\sigma=0.262$, $\Delta (B-V) = -0.028\pm0.035$ with $\sigma=0.149$, and $\Delta (V-I) = -0.032\pm0.023$ with $\sigma=0.234$ (this study minus Zloczewski \& Kaluzny 2009).}
\end{figure*}

\begin{figure*}
\centerline{
\includegraphics[height=140mm,angle=-90]{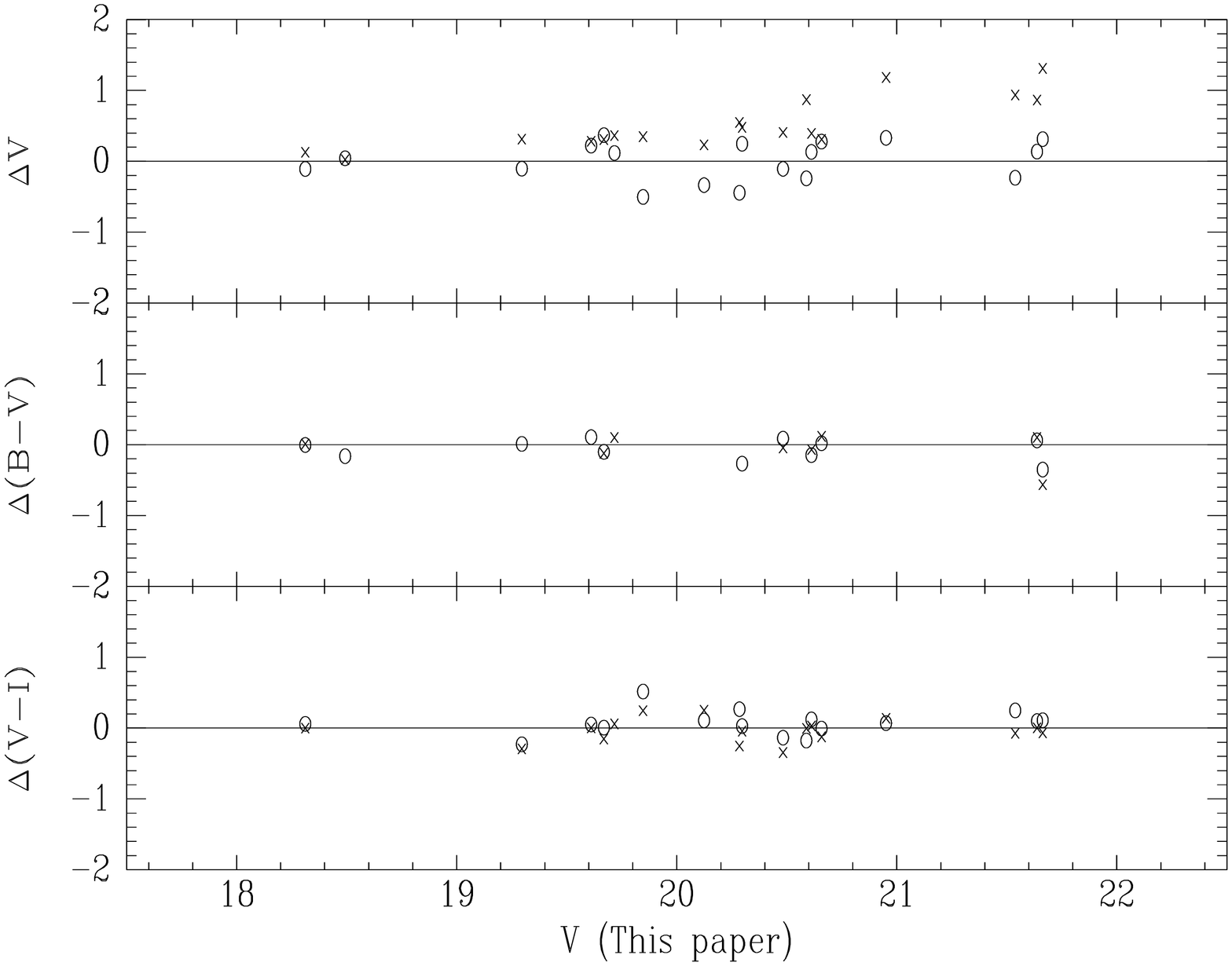}}
\vspace{0.5cm}
\caption{Comparisons of our photometry of M33 star clusters in the $UBVRI$ bands with previous measurements in \citet{Roman09} and \citet{ZK09}. Crosses and circles represent the photometry difference between this study and previous study of \citet{Roman09}, and the photometry difference between this study and previous study of \citet{ZK09}, respectively.  The photometric offsets and rms scatter of the differences between their measurements and our new magnitudes are: $\Delta V = 0.517\pm0.086$ with $\sigma=0.353$, $\Delta (B-V) = -0.050\pm0.079$ with $\sigma=0.208$, and $\Delta (V-I) = -0.037\pm0.040$ with $\sigma=0.161$ (this study minus San Roman et al. 2009); $\Delta V = 0.006\pm0.065$ with $\sigma=0.268$, $\Delta (B-V) = -0.068\pm0.045$ with $\sigma=0.143$, and $\Delta (V-I) = 0.071\pm0.045$ with $\sigma=0.173$ (this study minus Zloczewski \& Kaluzny 2009).}
\end{figure*}

From Figures 3--5, we can see that our measurements in the $V$ band get systematically fainter than the photometric measurements in \citet{Roman09} for fainter sources ($V\geq19$ mag). The $(V-I)$ colors obtained here are in good agreement with those in \citet{PL07} and \citet{Roman09}, however, the difference of $(B-V)$ colors between \citet{Roman09} and this paper is large, which turned out to be $0.388\pm 0.040$ with $\sigma=0.268$. From Figure 5, we can see that both the $(B-V)$ and $(V-I)$ colors obtained here are in good agreement with those in \citet{ZK09}, however, the $V$ difference between this study and \citet{ZK09} turned out to be $-0.103\pm 0.026$ with $\sigma=0.262$. By cross-identification, \citet{Roman09} provided 21 common star clusters in \citet{ZK09}. We derived photometry for 18 of these 21 star clusters. We compared the photometry of these 18 star clusters with previous measurements in \citet{Roman09} and \citet{ZK09} for comparison. Figure 6 shows the comparison. From Figure 6, we can see that our measurements in $V$ band get systematically fainter than the photometric measurements in \citet{Roman09} for fainter sources ($V\geq19$ mag), however, this trend disappears between this study and \citet{ZK09}. Both the $(B-V)$ and $(V-I)$ colors obtained here are in good agreement with those in \citet{Roman09} and \citet{ZK09}. In Paper I, we has discussed the $V$ difference between his study and previous studies in detail, and showed that the $V$ difference resulted from different photometric apertures adopted in his study and previous studies. In Paper I, we showed that if the photometric apertures were adopted in our study to be the same as previous studies, the $V$ difference disappeared.

\section{RESULTS}

In Paper I, we has presented some results for M33 star clusters including the CMD and color--color diagram. In addition, in Paper I, we pointed out that, before \citet{ZK09}, none of M33 star clusters with $V > 21.0$ mag has been detected. And \citet{ZK09} emphasized that the faintest known globular cluster in the Milky Way has $M_V \sim -1$ mag comparing with $M_V \sim -4$ mag ($V \sim 21$ mag) observed for the faintest of the known M33 globular cluster candidates before \citet{ZK09}. \citet{ZK09} provided integrated magnitudes for 115 M33 star clusters using the ACS/WFC images, of which nine have ${\rm 21.0~mag < V < 22.0~mag}$ corresponding to ${\rm -4~mag < M_V < -3~mag}$. Although the faintest star cluster of M33 detected by \citet{ZK09} is 2.0 mag brighter than the faintest Galactic globular cluster, it will provide something unique to the analysis of M33 star clusters when including them. In fact, Paper I included the photometry of the M33 star clusters in \citet{ZK09} when we provided the results for M33 star clusters, however, most star clusters in \citet{ZK09} have photometry in only two bands ($V$ and $I$). There are only 19 sample star clusters of \citet{ZK09} in the color--color diagram provided in Paper I. In addition, most star clusters in \citet{Roman09} also have photometry in only two bands ($V$ and $I$). So, it is necessary that we re-provide a CMD and color--color diagram of M33 including photometry obtained in this paper.

\begin{figure*}
\centerline{
\includegraphics[scale=0.8,angle=-90]{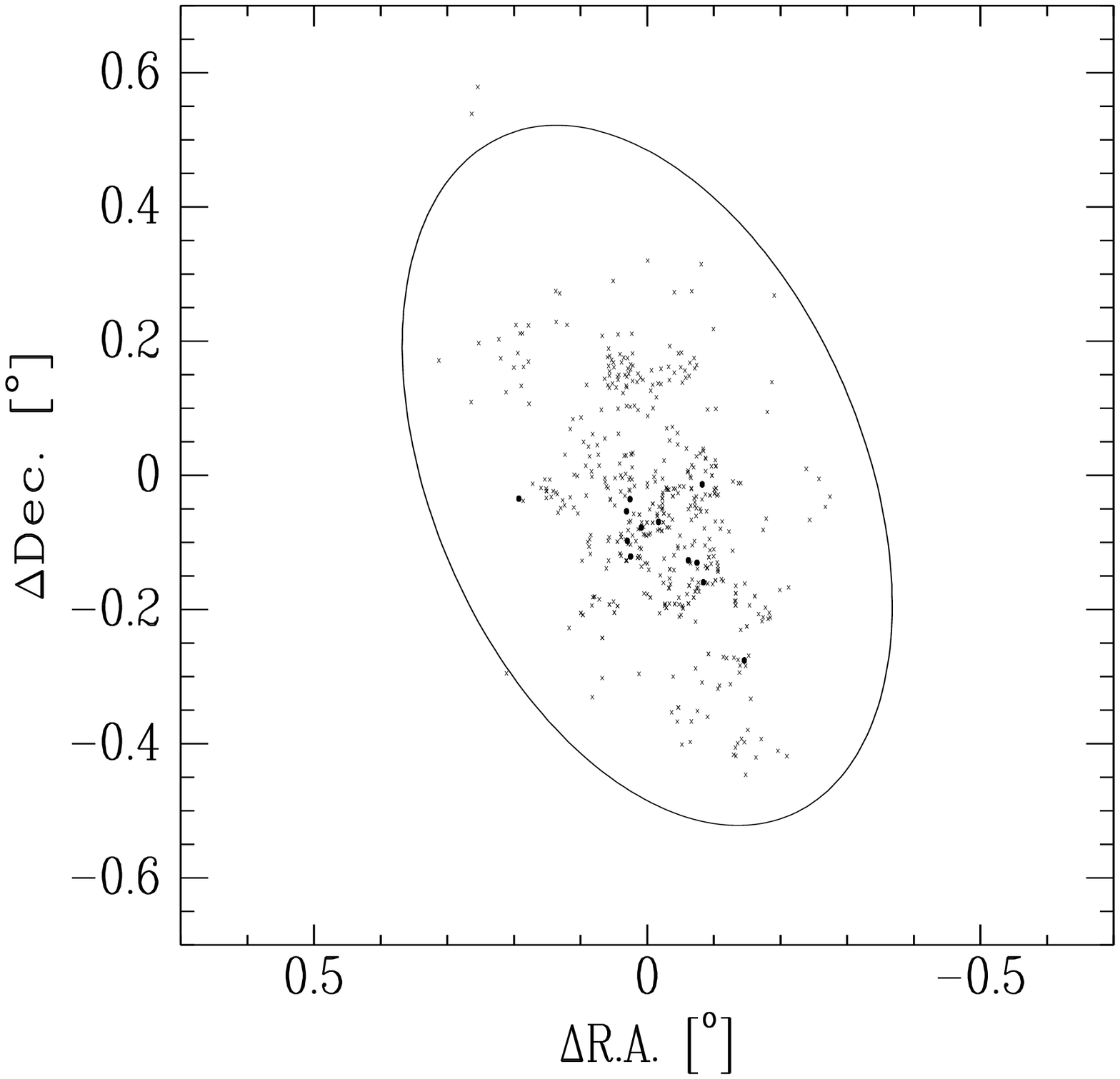}}
\caption[]{Spatial distribution of the 523 star clusters in M33. Crosses denote the star clusters, of which the photometry is obtained in Paper I and this study, and filled circles denote the star clusters, of which the photometry was obtained by \citet{PL07}, \citet{Roman09}, and \citet{ZK09}. The large ellipse is the $D_{25}$ boundary of the M33 disk \citep{Vaucouleurs91}.}
\label{fig10}
\end{figure*}

\subsection{Color--Magnitude Diagram}

The CMD can provide a qualitative model-independent global indication of cluster-formation history that can be compared between galaxies because $(B-V)_0$ and $(V-I)_0$ are reasonably good age indicators, at least between young and old populations, with a secondary dependence on metallicity \citep{CBF99b}. CMDs of M33 star clusters have been previously discussed in the literature (Christian \& Schommer 1982, 1988; Chandar et al. 1999b; Park \& Lee 2007; Paper I). However, with a much larger star cluster sample in this paper, it is worth investigating them again. This paper includes 523 star clusters of M33, of which the photometry of 234 and 277 is derived in this paper and in Paper I, respectively; and the photometry of the remaining 12 star clusters is from \citet{PL07}, \citet{Roman09} and \citet{ZK09}, since we cannot accurately derive the photometry for these 12 star clusters (see Section 2.2 for details). The 277 star clusters from Paper I are confirmed by \citet{sara07} (254 star clusters), \citet{PL07} (7 star clusters), and \citet{Roman09} (16 star clusters) based on the {\sl HST} and high-resolution ground-based imaging.

We point out that the photometry of M33 star clusters obtained in Paper I and this study is homogeneous photometric data in the same photometric system. For completeness of data and readers' references, we list the photometry of 277 star clusters of Paper I in Table 3 including the reddening values from \citet{PL07} and \citet{Roman09} in column 9 of Table 3 (Table 3 includes $E(B-V)$ missed in Paper I.). In Table 3, $R_C$ and $I_C$ mean that $RI$ magnitudes are on Johnson--Kron--Cousins system. For the reddening values of the star clusters, we used those from \citet{PL07} or \citet{Roman09}. For the star clusters, \citet{PL07} and \citet{Roman09} both presented their reddening values, we adopted their mean values. For the star clusters, \citet{PL07} and \citet{Roman09} did not present their reddening values, we adopted a uniform value of $E(B-V)=0.1$ typical of the published values for the line-of-sight reddenings to M33 that \citet{sara07} adopted. Figure 7 shows the spatial distribution of these 523 star clusters. The large ellipse is the $D_{25}$ boundary of the M33 disk \citep{Vaucouleurs91}. Figure 8 displays the integrated $M_V-(B-V)_0$ and $M_V-(V-I)_0$ CMDs of the sample star clusters of M33. The absolute magnitudes of the star clusters were derived for the adopted distance modulus of $(m-M)_0 = 24.64$ obtained by \citet{Galleti04}. The interstellar extinction curve, $A_{\lambda}$, is taken from \citet{Schlegel98}. Below each CMD in Figure 8 we plotted the star cluster distribution in color space. To the right of each CMD in Figure 8 we showed a histogram of the star clusters' absolute $V$ magnitudes.

\citet{sara07}, \citet{PL07}, and Paper I showed that the M33 star clusters are roughly separated into blue and red groups with a color boundary of $(B-V)_0\simeq0.5$ in the $M_V-(B-V)_0$ based on a small star cluster sample. However, this feature did not clearly appear in Figure 8 as previous studies (Sarajedini \& Mancone 2007; Park \& Lee 2007; Paper I). Figure 8 shows that the star cluster luminosity function peaks near $M_V\sim -6.0$ mag, and nearly half of star clusters lies between $M_V = -5.5$ and $M_V = -7.0$ mag.

By adding models to the CMDs, we can obtain a more detailed history of star cluster formation. Three fading lines ($M_V$ as a function of age) of \citet{bc03} for a metallicity of $Z=0.004, Y=0.24$ which are thought to be appropriate for M33 star clusters \citep{CBF99b}, assuming a Salpeter initial mass function (IMF; Salpeter 1955) with lower and upper-mass cut-offs of $m_{\rm L}=0.1~M_{\odot}$ and $m_{\rm U}=100~M_{\odot}$, and using the Padova-1994 evolutionary tracks, are plotted on the CMDs of
M33 star clusters for three different total initial masses: $10^5$, $10^4$, and $10^3$~$M_{\odot}$. The majority of M33 star clusters fall between these three fading lines. From Figure 8, we note that none of the youngest clusters ($\sim 10^7$~yr) have masses approaching $10^5$~$M_{\odot}$, which is consistent with the results of \citet{CBF99b} and Paper I. For ages older than $10^9$ yr, some clusters with substantially higher masses are seen.

\begin{figure*}
\centering
\includegraphics[height=135mm,angle=-90]{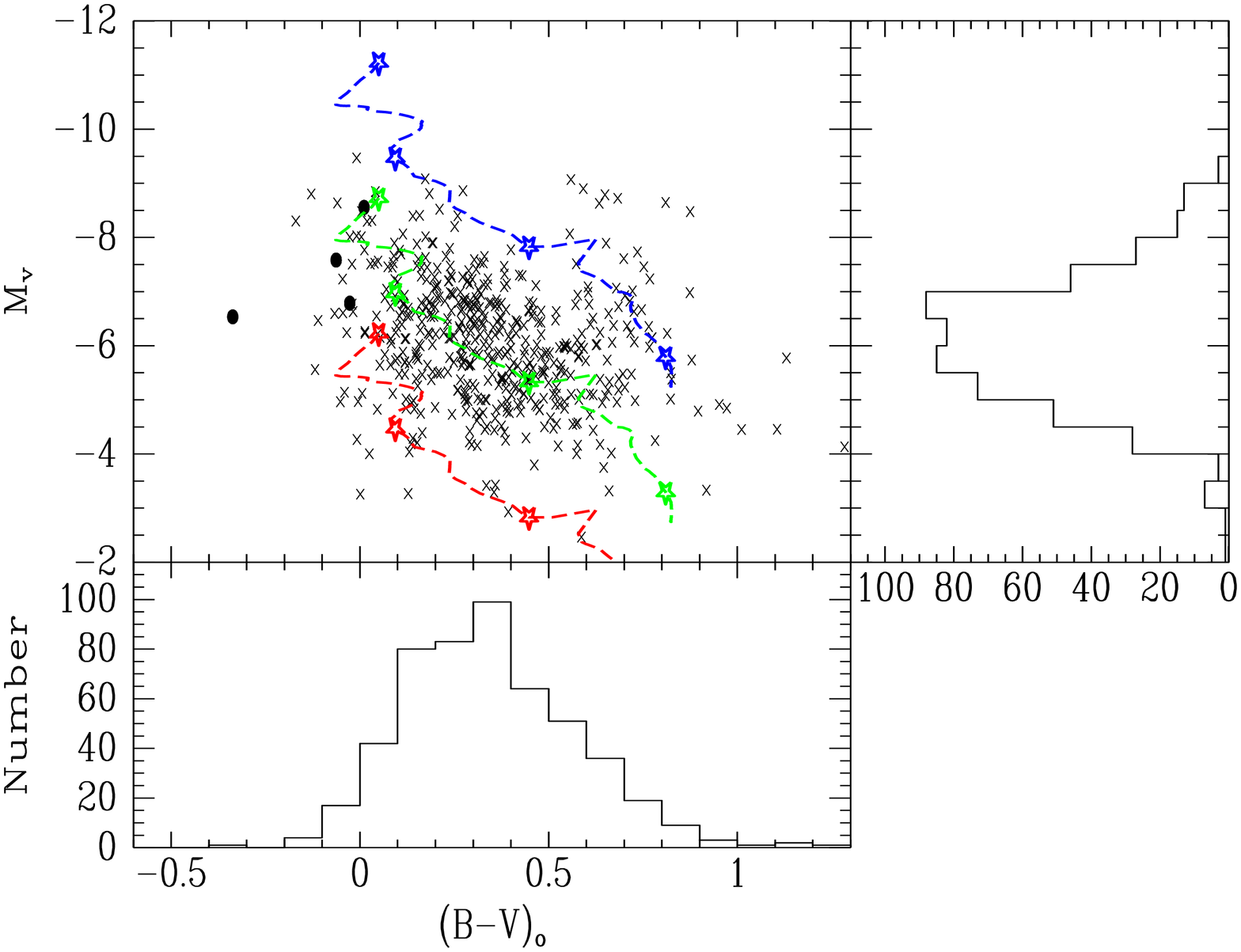}
\includegraphics[height=135mm,angle=-90]{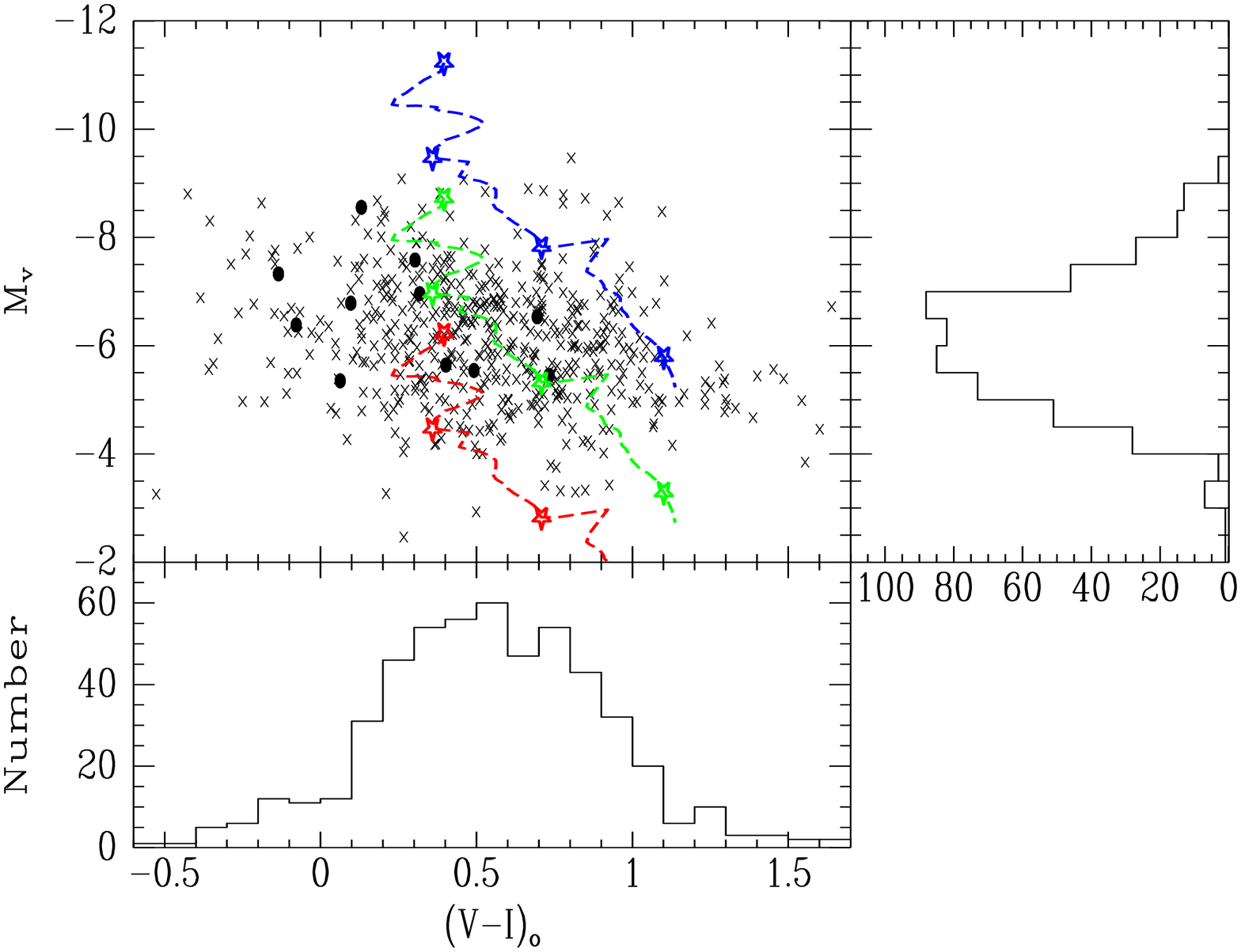}\\
\caption[]{Color--magnitude diagrams of M33 clusters. Crosses represent the star clusters in \citet{Ma12} and this study, filled circles represent the star clusters in
\citet{PL07}, \citet{Roman09}, and \citet{ZK09}. Fading lines are indicated for star clusters with total initial masses of $10^5$ (upper dashed line), $10^4$, and $10^3$ (lower dashed line) $M_\odot$, assuming a Salpeter IMF (see text). Stars along each fading line represent ages of $10^7$, $10^8$, $10^9$, and $10^{10}$ yr, from top to bottom, respectively.}
\end{figure*}

\subsection{Color--Color Diagram}

\begin{figure*}
\centerline{
\includegraphics[height=180mm,angle=-90]{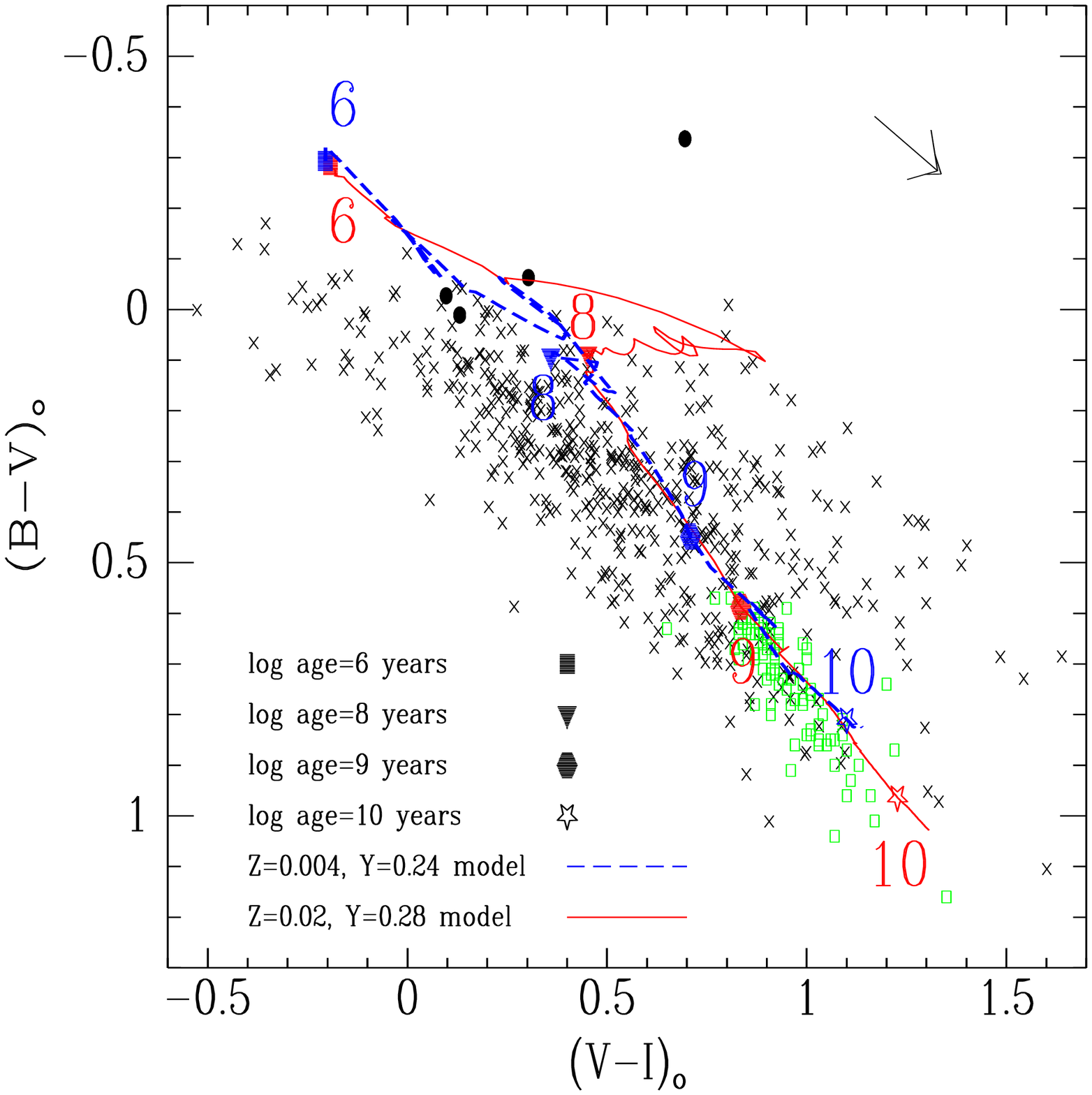}}
\caption{$(B-V)_0$ vs. $(V-I)_0$ color--color diagram of star clusters in M33. Crosses represent the star clusters in \citet{Ma12} and this study, filled circles represent the star clusters in \citet{PL07}, \citet{Roman09}, and \citet{ZK09}. Green squares are Galactic globular clusters from the online database of Harris (1996; 2010 update). Theoretical evolutionary paths from the SSP model (Bruzual \& Charlot 2003) for $Z=0.004$, $Y=0.24$ (blue dashed line) and $Z=0.02$, $Y=0.28$ (red solid line) are drawn for every dex in age from $10^6$ to $10^{10}$ yr. The arrow represents the reddening direction.}
\end{figure*}

Figure 9 shows the integrated $(B-V)_0$ versus $(V-I)_0$ color--color diagram for M33 star clusters. Galactic globular clusters from the online database of
Harris (1996; 2010 update) are also plotted for comparison. We overplotted the theoretical evolutionary path for the single stellar population (SSP; Bruzual \& Charlot 2003) for $Z=0.004, Y=0.24$ that was appropriate for M33 \citep{CBF99b}. To identify different time periods, the different symbols correspond to $10^6$, $10^7$, $10^8$, $10^9$, and $10^{10}$ yr. For comparison, the evolutionary path of the SSP for $Z=0.02, Y=0.28$ is also overlaid.

In general, the star clusters in M33 are located along the sequence that is consistent with the theoretical evolutionary path for $Z=0.004, Y=0.24$, while
some are on the redder or bluer side in the $(V-I)_0$ color. The wide color range of M33 star clusters implies a large range of ages, suggesting a prolonged epoch of formation. From Figure 9, we find that the photometry is shifted below the SSP lines, i.e., the sample star clusters are on the redder side in the $(B-V)_0$ color, when the star clusters have the $(V-I)_0$ color between $-0.5$ and $0.4$. In the same time, from Figure 9, we also find that the photometry for most of the Galactic globular clusters is also below the SSP lines but with much smaller range. Large scatter observed for M33 star clusters possibly results from large errors of colors. By comparing with SSP models, we can see that there are a large range of ages of M33 star clusters, of which some star clusters are as old as the Galactic globular clusters.

\section{SUMMARIES AND CONCLUSIONS}

In this paper, we present $UBVRI$ photometric measurements for 234 star clusters in the field of M33. These sample star clusters of M33 are from \citet{PL07}, \citet{Roman09} and \citet{ZK09}. For most of these star clusters, there is photometry in only two bands ($V$ and $I$) in previous studies. Photometry of these star clusters is performed using archival images from the LGGS \citep{massey}. Detailed comparisons show that, in general, our photometry is consistent with previous measurements, especially, our photometry is in good agreement with that of \citet{ZK09}. Combined with the star clusters' photometry in previous studies, we present some results:

1. None of the M33 youngest clusters ($\sim 10^7$~yr) have masses approaching $10^5$~$M_{\odot}$.

2. The wide color range of M33 star clusters implies a large range of ages, suggesting a prolonged epoch of formation. And comparisons with SSP models suggest a large range of ages of M33 star clusters, and some as old as the Galactic globular clusters.

\acknowledgments
We would like to thank the anonymous referee for providing rapid and thoughtful report that helped improve the original manuscript greatly. This research was supported by the Chinese National Natural Science Foundation through grants 10873016 and 10633020, and by National Basic Research Program of China (973 Program) under grant 2007CB815403.

\begin{sidewaystable}
\centering
\tiny
\caption{New $UBVRI$ Photometry of 234 M33 Star Clusters}
%\vspace{5mm}
\label{t1.tab}
\begin{tabular}{cccccccccccccc}
\tableline
\tableline
ID & ID & ID & ID & ID & R.A.  & Decl.  & $U$  & $B$ &$V$  & $R_C$   & $I_C$  & $E(B-V)$ & $r_{\rm ap}$ \\
 &  &  &  &  & (J2000.0) & (J2000.0) & (mag)  & (mag) & (mag) & (mag) & (mag) & (mag) & (\arcs) \\
\hline
  1...... &	    PL1 &         &         &         &	  01~32~36.337 & 30~36~49.41 &	 18.751 $\pm$  0.010 & 19.887 $\pm$  0.029 & 19.588 $\pm$  0.029 & 19.076 $\pm$  0.022 & 19.118 $\pm$  0.029 &   0.20 &  2.838\\
  2...... &	    PL2 &         &         &         &	  01~33~11.809 & 30~38~57.03 &	 16.677 $\pm$  0.017 & 17.536 $\pm$  0.018 & 17.326 $\pm$  0.018 & 17.129 $\pm$  0.018 & 17.429 $\pm$  0.020 &   0.10 &  2.064\\
  3...... &	    PL3 &         &         &         &	  01~33~12.732 & 30~38~55.69 &	 18.556 $\pm$  0.016 & 19.075 $\pm$  0.021 & 18.894 $\pm$  0.024 & 18.777 $\pm$  0.024 & 18.472 $\pm$  0.023 &   0.10 &  2.064\\
  4...... &	    PL4 &         &         &         &	  01~33~15.038 & 30~39~04.94 &	 18.604 $\pm$  0.017 & 18.918 $\pm$  0.019 & 18.463 $\pm$  0.018 & 18.092 $\pm$  0.018 & 17.700 $\pm$  0.017 &   0.15 &  3.354\\
  5...... &	    PL5 &         &         &         &	  01~33~19.558 & 30~35~30.00 &	 17.135 $\pm$  0.006 & 17.409 $\pm$  0.007 & 17.118 $\pm$  0.006 & 16.951 $\pm$  0.006 & 16.673 $\pm$  0.006 &   0.10 &  2.322\\
  6...... &	    PL6 &         &         &         &	  01~33~20.274 & 30~34~49.96 &	 17.246 $\pm$  0.006 & 17.756 $\pm$  0.011 & 17.570 $\pm$  0.011 & 17.531 $\pm$  0.014 & 17.308 $\pm$  0.018 &   0.10 &  4.128\\
\end{tabular}
\end{sidewaystable}
\addtocounter{table}{-1}

\begin{sidewaystable}
\centering
\tiny
\caption{(Continued.)}
%\vspace{5mm}
\label{t1.tab}
\begin{tabular}{cccccccccccccc}
\tableline
\tableline
ID & ID & ID & ID & ID & R.A.  & Decl.  & $U$  & $B$ &$V$  & $R_C$   & $I_C$  & $E(B-V)$ & $r_{\rm ap}$ \\
 &  &  &  &  & (J2000.0) & (J2000.0) & (mag)  & (mag) & (mag) & (mag) & (mag) & (mag) & (\arcs) \\
\hline
 57...... &	   SR31 &         &         &         &	  01~33~21.347 & 30~31~01.10 &	 20.644 $\pm$  0.068 & 20.353 $\pm$  0.056 & 19.354 $\pm$  0.028 & 18.857 $\pm$  0.024 & 18.194 $\pm$  0.018 &   0.12 &  3.096\\
 58...... &	   SR32 &         &         &         &	  01~33~21.375 & 30~31~32.05 &	 18.167 $\pm$  0.011 & 19.081 $\pm$  0.020 & 18.900 $\pm$  0.022 & 18.711 $\pm$  0.023 & 18.371 $\pm$  0.022 &   0.10 &  2.322\\
 59...... &	   SR34 &   SM62  &         &         &	  01~33~21.579 & 30~31~51.05 &	 19.093 $\pm$  0.023 & 19.231 $\pm$  0.024 & 18.515 $\pm$  0.018 & 18.005 $\pm$  0.014 & 17.468 $\pm$  0.011 &   0.10 &  2.838\\
 60...... &	   SR37 &         &         &         &	  01~33~22.215 & 30~38~27.90 &	 19.608 $\pm$  0.033 & 19.881 $\pm$  0.045 & 19.342 $\pm$  0.038 & 18.948 $\pm$  0.036 & 18.462 $\pm$  0.033 &   0.05 &  2.838\\
 61...... &	   SR38 &   SM69  &         &         &	  01~33~22.386 & 30~30~14.40 &	 17.117 $\pm$  0.007 & 17.483 $\pm$  0.009 & 17.205 $\pm$  0.009 & 17.075 $\pm$  0.011 & 16.729 $\pm$  0.011 &   0.05 &  4.644\\
 62...... &	   SR39 &         &         &         &	  01~33~22.746 & 30~38~00.01 &	 20.597 $\pm$  0.063 & 20.478 $\pm$  0.064 & 19.848 $\pm$  0.058 & 19.246 $\pm$  0.045 & 18.480 $\pm$  0.032 &   0.05 &  2.838\\
\end{tabular}
\end{sidewaystable}
\addtocounter{table}{-1}

\begin{sidewaystable}
\centering
\tiny
\caption{(Continued.)}
%\vspace{5mm}
\label{t1.tab}
\begin{tabular}{cccccccccccccc}
\tableline
\tableline
ID & ID & ID & ID & ID & R.A.  & Decl.  & $U$  & $B$ &$V$  & $R_C$   & $I_C$  & $E(B-V)$ & $r_{\rm ap}$ \\
 &  &  &  &  & (J2000.0) & (J2000.0) & (mag)  & (mag) & (mag) & (mag) & (mag) & (mag) & (\arcs) \\
\hline
113...... &	  SR113 &         &         &         &	  01~33~45.999 & 30~36~49.63 &	 18.000 $\pm$  0.017 & 18.775 $\pm$  0.017 & 18.597 $\pm$  0.021 & 18.549 $\pm$  0.033 & 18.461 $\pm$  0.040 &   0.08 &  2.064\\
114...... &	  SR115 &         &         &         &	  01~33~46.790 & 30~35~59.27 &	 20.428 $\pm$  0.029 & 20.539 $\pm$  0.031 & 19.985 $\pm$  0.029 & 19.549 $\pm$  0.029 & 19.056 $\pm$  0.024 &   0.12 &  1.548\\
115...... &	  SR116 &         &         &         &	  01~33~47.436 & 30~39~59.21 &	 20.204 $\pm$  0.054 & 20.022 $\pm$  0.043 & 19.566 $\pm$  0.046 & 19.425 $\pm$  0.064 & 19.400 $\pm$  0.133 &   0.08 &  1.548\\
116...... &	  SR117 &         &         &         &	  01~33~47.530 & 30~39~03.58 &	 18.437 $\pm$  0.031 & 18.709 $\pm$  0.042 & 18.329 $\pm$  0.048 & 18.117 $\pm$  0.054 & 17.804 $\pm$  0.065 &   0.05 &  2.838\\
117...... &	  SR118 &         &         &         &	  01~33~48.139 & 30~39~28.85 &	 19.589 $\pm$  0.046 & 20.117 $\pm$  0.066 & 19.533 $\pm$  0.066 & 19.068 $\pm$  0.067 & 18.425 $\pm$  0.063 &   0.10 &  1.290\\
118...... &	  SR119 &         &         &         &	  01~33~48.590 & 30~37~05.48 &	 18.387 $\pm$  0.011 & 18.927 $\pm$  0.015 & 18.430 $\pm$  0.015 & 18.080 $\pm$  0.015 & 17.593 $\pm$  0.014 &   0.19 &  2.064\\
\end{tabular}
\end{sidewaystable}
\addtocounter{table}{-1}

\begin{sidewaystable}
\centering
\tiny
\caption{(Continued.)}
%\vspace{5mm}
\label{t1.tab}
\begin{tabular}{cccccccccccccc}
\tableline
\tableline
ID & ID & ID & ID & ID & R.A.  & Decl.  & $U$  & $B$ &$V$  & $R_C$   & $I_C$  & $E(B-V)$ & $r_{\rm ap}$ \\
 &  &  &  &  & (J2000.0) & (J2000.0) & (mag)  & (mag) & (mag) & (mag) & (mag) & (mag) & (\arcs) \\
\hline
169...... &	  ZK-23 &         &         &         &	  01~33~30.847 & 30~29~14.41 &	 18.260 $\pm$  0.009 & 18.607 $\pm$  0.013 & 18.396 $\pm$  0.013 & 18.209 $\pm$  0.013 & 17.889 $\pm$  0.013 &   0.10 &  2.322\\
170...... &	  ZK-24 &         &         &         &	  01~33~31.510 & 30~28~42.08 &	 18.953 $\pm$  0.014 & 18.933 $\pm$  0.016 & 18.541 $\pm$  0.015 & 18.304 $\pm$  0.015 & 17.796 $\pm$  0.013 &   0.10 &  2.580\\
171...... &	  ZK-25 &         &         &         &	  01~33~32.378 & 30~17~36.35 &	 22.222 $\pm$  0.060 & 22.181 $\pm$  0.065 & 21.725 $\pm$  0.062 & 21.403 $\pm$  0.066 & 20.771 $\pm$  0.061 &   0.10 &  1.548\\
172...... &	  ZK-26 &         &         &         &	  01~33~33.436 & 30~29~45.80 &	 19.001 $\pm$  0.012 & 19.535 $\pm$  0.021 & 19.381 $\pm$  0.025 & 19.105 $\pm$  0.027 & 18.447 $\pm$  0.022 &   0.10 &  2.322\\
173...... &	  ZK-27 &         &         &         &	  01~33~33.696 & 30~28~09.65 &	 18.234 $\pm$  0.011 & 18.558 $\pm$  0.014 & 18.331 $\pm$  0.014 & 18.354 $\pm$  0.021 & 18.287 $\pm$  0.028 &   0.10 &  2.838\\
174...... &	  ZK-29 &         &         &         &	  01~33~33.959 & 30~39~53.96 &	 19.079 $\pm$  0.019 & 19.342 $\pm$  0.024 & 18.847 $\pm$  0.024 & 18.492 $\pm$  0.022 & 17.985 $\pm$  0.026 &   0.10 &  2.580\\
\end{tabular}
\end{sidewaystable}
\addtocounter{table}{-1}

\begin{sidewaystable}
\centering
\tiny
\caption{Continued.}
%\vspace{5mm}
\label{t1.tab}
\begin{tabular}{cccccccccccccc}
\tableline
\tableline
ID & ID & ID & ID & ID & R.A.  & Decl.  & $U$  & $B$ &$V$  & $R_C$   & $I_C$  & $E(B-V)$ & $r_{\rm ap}$ \\
 &  &  &  &  & (J2000.0) & (J2000.0) & (mag)  & (mag) & (mag) & (mag) & (mag) & (mag) & (\arcs) \\
\hline
225...... &	  ZK-82 &         &         &         &	  01~34~06.750 & 30~34~05.20 &	      ......         & 20.864 $\pm$  0.054 & 20.036 $\pm$  0.044 & 19.278 $\pm$  0.029 & 18.355 $\pm$  0.017 &   0.10 &  1.806\\
226...... &	  ZK-83 &         &         &         &	  01~34~07.277 & 30~33~46.31 &	 20.171 $\pm$  0.034 & 20.105 $\pm$  0.033 & 19.472 $\pm$  0.034 & 19.104 $\pm$  0.036 & 18.675 $\pm$  0.046 &   0.10 &  3.354\\
227...... &	  ZK-84 &         &         &         &	  01~34~09.691 & 30~25~05.00 &	 19.821 $\pm$  0.020 & 20.094 $\pm$  0.023 & 19.907 $\pm$  0.028 & 19.703 $\pm$  0.030 & 19.589 $\pm$  0.039 &   0.10 &  2.580\\
228...... &	  ZK-85 &         &         &         &	  01~34~10.933 & 30~28~33.04 &	 19.109 $\pm$  0.013 & 19.553 $\pm$  0.018 & 19.334 $\pm$  0.023 & 19.341 $\pm$  0.033 & 19.356 $\pm$  0.043 &   0.10 &  3.096\\
229...... &	  ZK-86 &         &         &         &	  01~34~12.933 & 30~28~46.25 &	 18.094 $\pm$  0.007 & 18.432 $\pm$  0.009 & 18.121 $\pm$  0.009 & 17.939 $\pm$  0.009 & 17.683 $\pm$  0.009 &   0.10 &  2.322\\
230...... &	  ZK-87 &         &         &         &	  01~34~13.533 & 30~28~44.05 &	 20.513 $\pm$  0.030 & 20.529 $\pm$  0.031 & 19.916 $\pm$  0.032 & 19.479 $\pm$  0.031 & 18.814 $\pm$  0.028 &   0.10 &  2.580\\
\end{tabular}
\end{sidewaystable}
%\enddata
%\end{deluxetable}

\begin{table}
\begin{center}
\caption{Comparison between this Study and Previous studies of $V$ Photometry for M33 Star Clusters Considered Here}
%\vspace{5mm}
\label{t2.tab}
\begin{tabular}{cccccccc}
\tableline
\tableline
%\tabletypesize{\scriptsize}
${{\rm ID}^{a}}$ & ${{\rm ID}^{b}}$ & ${V^{c}}$ & ${V^{d}}$ & ${V^{e}}$ & ${\Delta V^{f}}$ & ${\Delta V^{g}}$ &
${r_{\rm ap}^{h}}$ \\
 &  & (mag) & (mag) & (mag) & (mag) & (mag) & (\arcs)\\
\hline
   PL197 &        &17.652$\pm$0.003 &       & 20.384$\pm$0.056 & 2.732 &       & 1.032\\
   SR5   & 25-1-003 & 19.768           & 20.41 & 20.951$\pm$0.087 & 1.183 & 0.541 & 1.290\\
   SR16  &33-5-019&20.352           & 21.16 & 21.663$\pm$0.113 & 1.311 & 0.503 & 2.580\\
   SR45  &        &20.063           &       & 21.159$\pm$0.060 & 1.095 &       &1.548\\
   SR50  &        &19.530           &       & 20.670$\pm$0.174 & 1.140 &       &2.580\\
   SR58  &        &19.514           &       & 20.601$\pm$0.047 & 1.087 &       &1.548\\
   SR116 &        &18.405           &       & 19.566$\pm$0.046 & 1.161 &       &1.548\\
   SR118 &        &18.151           &       & 19.533$\pm$0.066 & 1.382 &       &1.290\\
\hline
\end{tabular}
\end{center}
{$^a$The star cluster names following the naming convention of \citet{PL07} or \citet{Roman09}.}\\
$^{b}${The star cluster names following the naming convention of \citet{ZK09}.}\\
$^{c}${The photometry obtained by \citet{PL07} or by \citet{Roman09}.}\\
$^{d}${The photometry obtained by \citet{ZK09}.}\\
$^{e}${The photometry obtained in this paper.}\\
$^{f}${The magnitude difference between this study and \citet{PL07} or \citet{Roman09} (this study minus Park \& Lee 2007 or San Roman et al. 2009).}\\
$^{g}${The magnitude difference between this study and \citet{ZK09} (this study minus
Zloczewski \& Kaluzny 2009).}\\
$^{h}${The aperture radius of photometry adopted in this paper.}\\
\end{table}

\begin{sidewaystable}
\centering
\tiny
\caption{$UBVRI$ Photometry of 277 M33 Star Clusters in \citet{Ma12}}
%\vspace{5mm}
\label{t1.tab}
\begin{tabular}{ccccccccccccc}
\tableline
\tableline
ID & ID & ID &  ID & R.A.  & Decl.  & $U$  & $B$ &$V$  & $R_C$   & $I_C$  & $E(B-V)$ & $r_{\rm ap}$ \\
 &  &  &  &  (J2000.0) & (J2000.0) & (mag)  & (mag) & (mag) & (mag) & (mag) & (mag) & (\arcs) \\
\hline
235...... &	    SM3 &    PL33 &         &	  01~32~34.430 & 30~37~42.61 &	 21.184 $\pm$  0.053 & 21.087 $\pm$  0.064 & 20.419 $\pm$  0.042 & 20.107 $\pm$  0.041 & 19.807 $\pm$  0.037 &   0.05 & 2.322 \\
236...... &	    SM6 &   PL105 &         &	  01~32~38.977 & 30~39~18.03 &	 20.223 $\pm$  0.035 & 20.276 $\pm$  0.047 & 19.767 $\pm$  0.042 & 19.444 $\pm$  0.038 & 18.970 $\pm$  0.032 &   0.10 & 3.096 \\
237...... &	    SM9 &         &         &	  01~32~42.944 & 30~35~38.67 &	 17.912 $\pm$  0.007 & 18.016 $\pm$  0.009 & 17.609 $\pm$  0.008 & 17.331 $\pm$  0.007 & 16.929 $\pm$  0.006 &   0.10 & 3.612 \\
238...... &	   SM10 &   PL106 &         &	  01~32~44.302 & 30~40~12.33 &	 19.741 $\pm$  0.028 & 19.597 $\pm$  0.025 & 18.780 $\pm$  0.017 & 18.294 $\pm$  0.016 & 17.676 $\pm$  0.011 &   0.10 & 3.096 \\
239...... &	   SM15 &         &         &	  01~32~51.818 & 30~29~36.52 &	 18.937 $\pm$  0.015 & 18.985 $\pm$  0.020 & 18.512 $\pm$  0.015 & 18.178 $\pm$  0.013 & 17.724 $\pm$  0.012 &   0.10 & 3.354 \\
240...... &	   SM16 &         &         &	  01~32~52.644 & 30~14~30.97 &	 19.114 $\pm$  0.010 & 19.191 $\pm$  0.012 & 18.842 $\pm$  0.010 & 18.640 $\pm$  0.011 & 18.304 $\pm$  0.011 &   0.10 & 2.580 \\
241...... &	   SM23 &         &         &	  01~32~55.473 & 30~29~22.34 &	 19.436 $\pm$  0.017 & 19.765 $\pm$  0.026 & 19.545 $\pm$  0.025 & 19.280 $\pm$  0.024 & 18.666 $\pm$  0.018 &   0.10 & 2.064 \\
242...... &	   SM26 &         &         &	  01~32~56.323 & 30~14~58.94 &	 18.171 $\pm$  0.007 & 18.125 $\pm$  0.007 & 17.752 $\pm$  0.007 & 17.536 $\pm$  0.007 & 17.215 $\pm$  0.009 &   0.10 & 4.128 \\
243...... &	   SM28 &         &         &	  01~32~57.600 & 30~55~42.72 &	 20.833 $\pm$  0.035 & 20.597 $\pm$  0.032 & 19.840 $\pm$  0.025 & 19.351 $\pm$  0.021 & 18.948 $\pm$  0.021 &   0.10 & 2.838 \\
244...... &	   SM29 &         &         &	  01~32~58.626 & 30~47~57.23 &	 19.454 $\pm$  0.013 & 19.528 $\pm$  0.015 & 19.273 $\pm$  0.016 & 19.222 $\pm$  0.019 & 19.032 $\pm$  0.021 &   0.10 & 2.580 \\
245...... &	   SM31 &         &     SR3 &	  01~33~00.380 & 30~26~47.95 &	 20.538 $\pm$  0.042 & 20.619 $\pm$  0.052 & 20.070 $\pm$  0.037 & 19.696 $\pm$  0.031 & 19.241 $\pm$  0.024 &   0.08 & 2.322 \\
246...... &	   SM32 &         &         &	  01~33~00.542 & 30~45~17.60 &	 18.829 $\pm$  0.010 & 19.070 $\pm$  0.012 & 18.787 $\pm$  0.012 & 18.668 $\pm$  0.014 & 18.519 $\pm$  0.017 &   0.10 & 1.806 \\
247...... &	   SM34 &   PL107 &         &	  01~33~01.124 & 30~35~45.23 &	 20.661 $\pm$  0.037 & 20.895 $\pm$  0.052 & 20.356 $\pm$  0.046 & 19.783 $\pm$  0.036 & 19.220 $\pm$  0.027 &   0.05 & 1.806 \\
248...... &	   SM35 &   PL108 &         &	  01~33~02.366 & 30~34~44.42 &	 17.657 $\pm$  0.006 & 18.675 $\pm$  0.012 & 18.592 $\pm$  0.016 & 18.341 $\pm$  0.016 & 18.534 $\pm$  0.021 &   0.15 & 1.806 \\
249...... &	   SM40 &   PL109 &    SR11 &	  01~33~08.111 & 30~28~00.26 &	 19.872 $\pm$  0.033 & 19.815 $\pm$  0.036 & 19.027 $\pm$  0.025 & 18.674 $\pm$  0.023 & 18.244 $\pm$  0.019 &   0.12 & 3.096 \\
250...... &	   SM41 &         &         &	  01~33~09.818 & 30~12~50.70 &	 19.058 $\pm$  0.010 & 19.095 $\pm$  0.012 & 18.810 $\pm$  0.012 & 18.668 $\pm$  0.014 & 18.462 $\pm$  0.017 &   0.10 & 3.096 \\
251...... &	   SM42 &   PL110 &         &	  01~33~10.114 & 30~29~56.97 &	 18.324 $\pm$  0.030 & 18.689 $\pm$  0.027 & 18.557 $\pm$  0.026 & 18.408 $\pm$  0.033 & 18.175 $\pm$  0.031 &   0.15 & 3.612 \\
252...... &	   SM44 &         &         &	  01~33~13.804 & 30~29~03.60 &	 20.989 $\pm$  0.044 & 20.823 $\pm$  0.046 & 20.180 $\pm$  0.038 & 19.789 $\pm$  0.034 & 19.408 $\pm$  0.037 &   0.10 & 1.290 \\
253...... &	   SM45 &   PL111 &         &	  01~33~13.874 & 30~29~05.30 &	 21.442 $\pm$  0.055 & 21.055 $\pm$  0.046 & 20.170 $\pm$  0.032 & 19.610 $\pm$  0.024 & 19.120 $\pm$  0.023 &   0.20 & 1.290 \\
254...... &	   SM46 &         &         &	  01~33~13.887 & 30~28~24.33 &	 23.105 $\pm$  0.190 & 23.243 $\pm$  0.222 & 22.556 $\pm$  0.163 & 21.946 $\pm$  0.118 & 22.152 $\pm$  0.159 &   0.10 & 1.032 \\
255...... &	   SM47 &   PL112 &         &	  01~33~13.904 & 30~29~44.79 &	 17.549 $\pm$  0.014 & 18.519 $\pm$  0.020 & 18.303 $\pm$  0.020 & 18.216 $\pm$  0.021 & 18.482 $\pm$  0.029 &   0.15 & 2.580 \\
256...... &	   SM49 &   PL113 &    SR22 &	  01~33~14.330 & 30~28~22.90 &	 19.362 $\pm$  0.021 & 19.129 $\pm$  0.020 & 18.323 $\pm$  0.013 & 17.752 $\pm$  0.010 & 17.258 $\pm$  0.008 &   0.08 & 3.096 \\
257...... &	   SM52 &   PL114 &         &	  01~33~15.179 & 30~32~53.06 &	 20.467 $\pm$  0.041 & 20.479 $\pm$  0.044 & 19.778 $\pm$  0.035 & 19.410 $\pm$  0.031 & 19.018 $\pm$  0.029 &   0.05 & 2.322 \\
258...... &	   SM53 &         &    SR25 &	  01~33~16.089 & 30~20~56.69 &	 18.609 $\pm$  0.007 & 18.657 $\pm$  0.009 & 18.285 $\pm$  0.010 & 18.055 $\pm$  0.010 & 17.645 $\pm$  0.010 &   0.08 & 3.096 \\
259...... &	   SM54 &   PL115 &         &	  01~33~16.631 & 30~34~35.82 &	 20.327 $\pm$  0.031 & 20.227 $\pm$  0.033 & 19.683 $\pm$  0.026 & 19.208 $\pm$  0.025 & 18.812 $\pm$  0.023 &   0.10 & 2.064 \\
260...... &	   SM61 &         &    SR29 &	  01~33~21.169 & 30~37~55.61 &	 20.024 $\pm$  0.030 & 19.868 $\pm$  0.034 & 19.274 $\pm$  0.030 & 18.946 $\pm$  0.030 & 18.346 $\pm$  0.027 &   0.08 & 2.838 \\
261...... &	   SM63 &         &    SR35 &	  01~33~21.667 & 30~37~48.52 &	 20.059 $\pm$  0.025 & 20.092 $\pm$  0.029 & 19.569 $\pm$  0.030 & 19.245 $\pm$  0.031 & 18.945 $\pm$  0.031 &   0.12 & 2.064 \\
262...... &	   SM65 &         &         &	  01~33~22.146 & 30~45~34.33 &	 19.846 $\pm$  0.024 & 19.674 $\pm$  0.022 & 19.034 $\pm$  0.019 & 18.596 $\pm$  0.020 & 18.083 $\pm$  0.020 &   0.10 & 3.354 \\
263...... &	   SM66 &    PL34 &    SR36 &	  01~33~22.110 & 30~40~28.34 &	 20.577 $\pm$  0.035 & 20.479 $\pm$  0.042 & 19.249 $\pm$  0.026 & 18.307 $\pm$  0.014 & 17.348 $\pm$  0.012 &   0.10 & 1.548 \\
264...... &	   SM67 &         &         &	  01~33~22.167 & 30~40~25.97 &	 18.273 $\pm$  0.008 & 18.678 $\pm$  0.011 & 18.473 $\pm$  0.012 & 18.421 $\pm$  0.013 & 18.186 $\pm$  0.015 &   0.10 & 1.806 \\
265...... &	   SM68 &    PL35 &         &	  01~33~22.322 & 30~40~59.37 &	 18.834 $\pm$  0.017 & 18.962 $\pm$  0.018 & 18.630 $\pm$  0.022 & 18.498 $\pm$  0.026 & 18.006 $\pm$  0.026 &   0.10 & 3.096 \\
266...... &	   SM70 &         &         &	  01~33~23.112 & 30~33~00.59 &	 17.559 $\pm$  0.006 & 17.893 $\pm$  0.008 & 17.588 $\pm$  0.008 & 17.410 $\pm$  0.008 & 17.090 $\pm$  0.008 &   0.10 & 2.322 \\
267...... &	   SM71 &         &    SR41 &	  01~33~23.109 & 30~32~22.94 &	 19.768 $\pm$  0.027 & 19.769 $\pm$  0.031 & 19.121 $\pm$  0.021 & 18.725 $\pm$  0.018 & 18.300 $\pm$  0.018 &   0.12 & 2.580 \\
268...... &	   SM74 &    PL37 &    SR42 &	  01~33~23.905 & 30~40~26.05 &	 19.858 $\pm$  0.029 & 20.070 $\pm$  0.037 & 19.603 $\pm$  0.039 & 19.414 $\pm$  0.051 & 18.899 $\pm$  0.052 &   0.08 & 2.580 \\
269...... &	   SM75 &   PL116 &         &	  01~33~24.616 & 30~32~56.19 &	 20.682 $\pm$  0.041 & 20.538 $\pm$  0.043 & 19.917 $\pm$  0.034 & 19.551 $\pm$  0.030 & 18.916 $\pm$  0.027 &   0.10 & 2.064 \\
270...... &	   SM76 &    PL38 &         &	  01~33~24.856 & 30~33~55.07 &	 19.825 $\pm$  0.023 & 20.280 $\pm$  0.034 & 20.039 $\pm$  0.037 & 19.970 $\pm$  0.048 & 19.501 $\pm$  0.041 &   0.10 & 2.064 \\
271...... &	   SM77 &    PL39 &         &	  01~33~25.622 & 30~29~56.88 &	 19.432 $\pm$  0.018 & 19.203 $\pm$  0.018 & 18.356 $\pm$  0.011 & 17.828 $\pm$  0.009 & 17.282 $\pm$  0.007 &   0.10 & 2.322 \\
272...... &	   SM78 &         &         &	  01~33~25.594 & 30~45~30.67 &	 19.065 $\pm$  0.015 & 19.222 $\pm$  0.018 & 18.882 $\pm$  0.021 & 18.655 $\pm$  0.026 & 18.384 $\pm$  0.030 &   0.10 & 3.096 \\
273...... &	   SM80 &    PL40 &         &	  01~33~25.999 & 30~36~24.38 &	 18.128 $\pm$  0.012 & 18.197 $\pm$  0.012 & 17.811 $\pm$  0.011 & 17.596 $\pm$  0.013 & 17.278 $\pm$  0.014 &   0.10 & 2.064 \\
274...... &	   SM81 &    PL41 &         &	  01~33~26.371 & 30~41~06.89 &	 19.671 $\pm$  0.024 & 19.655 $\pm$  0.022 & 19.085 $\pm$  0.021 & 18.670 $\pm$  0.018 & 18.137 $\pm$  0.015 &   0.20 & 2.064 \\
275...... &	   SM83 &    PL42 &         &	  01~33~26.493 & 30~41~11.61 &	 20.234 $\pm$  0.043 & 20.186 $\pm$  0.037 & 19.696 $\pm$  0.034 & 19.486 $\pm$  0.036 & 19.032 $\pm$  0.034 &   0.20 & 2.064 \\
276...... &	   SM85 &    PL43 &         &	  01~33~26.765 & 30~33~21.43 &	 17.387 $\pm$  0.006 & 17.836 $\pm$  0.009 & 17.574 $\pm$  0.010 & 17.402 $\pm$  0.011 & 17.109 $\pm$  0.012 &   0.10 & 2.064 \\
277...... &	   SM86 &    PL44 &         &	  01~33~26.943 & 30~34~52.54 &	 18.774 $\pm$  0.012 & 19.074 $\pm$  0.018 & 18.648 $\pm$  0.018 & 18.285 $\pm$  0.020 & 17.687 $\pm$  0.018 &   0.10 & 2.838 \\
278...... &	   SM87 &    PL45 &         &	  01~33~27.370 & 30~41~59.89 &	 19.149 $\pm$  0.025 & 19.197 $\pm$  0.024 & 18.420 $\pm$  0.021 & 17.789 $\pm$  0.018 & 17.247 $\pm$  0.017 &   0.20 & 3.612 \\
279...... &	   SM88 &    PL46 &         &	  01~33~27.967 & 30~37~28.41 &	 19.292 $\pm$  0.023 & 19.707 $\pm$  0.032 & 19.463 $\pm$  0.034 & 19.194 $\pm$  0.029 & 18.971 $\pm$  0.032 &   0.15 & 1.806 \\
280...... &	   SM90 &   PL117 &         &	  01~33~28.009 & 30~21~06.19 &	 20.020 $\pm$  0.017 & 20.036 $\pm$  0.018 & 19.721 $\pm$  0.007 & 19.373 $\pm$  0.018 & 19.259 $\pm$  0.025 &   0.15 & 1.806 \\
281...... &	   SM91 &         &         &	  01~33~28.111 & 30~58~30.73 &	 18.940 $\pm$  0.022 & 18.657 $\pm$  0.016 & 17.788 $\pm$  0.011 & 17.236 $\pm$  0.008 & 16.692 $\pm$  0.008 &   0.10 & 4.128 \\
282...... &	   SM93 &    PL47 &         &	  01~33~28.414 & 30~36~23.08 &	 17.922 $\pm$  0.014 & 17.919 $\pm$  0.014 & 17.492 $\pm$  0.014 & 17.244 $\pm$  0.017 & 16.784 $\pm$  0.016 &   0.10 & 3.612 \\
283...... &	   SM94 &    PL48 &         &	  01~33~28.680 & 30~36~37.58 &	 17.922 $\pm$  0.010 & 18.052 $\pm$  0.011 & 17.711 $\pm$  0.011 & 17.500 $\pm$  0.012 & 17.198 $\pm$  0.013 &   0.10 & 2.580 \\
284...... &	   SM95 &    PL49 &         &	  01~33~28.719 & 30~41~35.13 &	 18.092 $\pm$  0.011 & 18.042 $\pm$  0.010 & 17.286 $\pm$  0.008 & 16.846 $\pm$  0.008 & 16.379 $\pm$  0.007 &   0.20 & 3.096 \\
285...... &	   SM98 &    PL50 &    SR55 &	  01~33~29.485 & 30~30~02.18 &	 18.044 $\pm$  0.009 & 18.387 $\pm$  0.012 & 18.120 $\pm$  0.012 & 17.975 $\pm$  0.015 & 17.689 $\pm$  0.016 &   0.09 & 2.580 \\
286...... &	  SM100 &   PL118 &         &	  01~33~30.070 & 30~49~29.18 &	 21.212 $\pm$  0.038 & 21.242 $\pm$  0.041 & 20.846 $\pm$  0.046 & 20.640 $\pm$  0.062 & 20.343 $\pm$  0.067 &   0.10 & 1.806 \\
287...... &	  SM101 &         &         &	  01~33~30.695 & 30~26~32.00 &	 19.138 $\pm$  0.023 & 18.832 $\pm$  0.021 & 18.010 $\pm$  0.013 & 17.518 $\pm$  0.011 & 16.917 $\pm$  0.009 &   0.10 & 3.612 \\
288...... &	  SM102 &   PL119 &         &	  01~33~30.698 & 30~22~21.55 &	 18.988 $\pm$  0.026 & 18.741 $\pm$  0.022 & 18.036 $\pm$  0.014 & 17.644 $\pm$  0.012 & 17.206 $\pm$  0.010 &   0.10 & 3.870 \\
289...... &	  SM103 &   PL120 &         &	  01~33~30.894 & 30~49~11.97 &	 19.216 $\pm$  0.016 & 19.034 $\pm$  0.015 & 18.478 $\pm$  0.014 & 18.142 $\pm$  0.017 & 17.722 $\pm$  0.015 &   0.10 & 3.870 \\
290...... &	  SM104 &    PL51 &         &	  01~33~30.924 & 30~37~12.90 &	 18.492 $\pm$  0.011 & 18.711 $\pm$  0.014 & 18.376 $\pm$  0.016 & 18.111 $\pm$  0.018 & 17.681 $\pm$  0.021 &   0.15 & 2.838 \\
291...... &	  SM105 &    PL52 &         &	  01~33~30.995 & 30~36~52.62 &	 17.983 $\pm$  0.010 & 18.380 $\pm$  0.015 & 18.040 $\pm$  0.016 & 17.752 $\pm$  0.016 & 17.125 $\pm$  0.014 &   0.10 & 2.580 \\
\end{tabular}
\end{sidewaystable}
\addtocounter{table}{-1}

\begin{sidewaystable}
\centering
\tiny
\caption{(Continued.)}
%\vspace{5mm}
\label{t3.tab}
\begin{tabular}{ccccccccccccc}
\tableline
\tableline
ID & ID & ID & ID  & R.A.  & Decl.  & $U$  & $B$ &$V$  & $R_C$   & $I_C$  & $E(B-V)$ & $r_{\rm ap}$ \\
 &  &  &  &  (J2000.0) & (J2000.0) & (mag)  & (mag) & (mag) & (mag) & (mag) & (mag) & (\arcs) \\
\hline
292...... &	  SM106 &   PL121 &         &	  01~33~31.103 & 30~33~45.55 &	 19.034 $\pm$  0.017 & 18.988 $\pm$  0.020 & 18.650 $\pm$  0.021 & 18.387 $\pm$  0.024 & 18.135 $\pm$  0.027 &   0.15 & 2.580 \\
293...... &	  SM109 &   PL122 &         &	  01~33~31.238 & 30~50~07.22 &	 19.799 $\pm$  0.025 & 19.766 $\pm$  0.025 & 19.372 $\pm$  0.029 & 19.083 $\pm$  0.030 & 18.775 $\pm$  0.033 &   0.10 & 3.354 \\
294...... &	  SM110 &    PL53 &    SR61 &	  01~33~31.406 & 30~40~20.37 &	 18.750 $\pm$  0.015 & 18.581 $\pm$  0.015 & 17.979 $\pm$  0.013 & 17.619 $\pm$  0.015 & 17.241 $\pm$  0.017 &   0.18 & 3.354 \\
295...... &	  SM113 &   PL123 &         &	  01~33~32.019 & 30~33~21.81 &	 17.160 $\pm$  0.006 & 17.610 $\pm$  0.008 & 17.309 $\pm$  0.008 & 17.099 $\pm$  0.008 & 16.750 $\pm$  0.009 &   0.15 & 2.322 \\
296...... &	  SM114 &    PL54 &    SR63 &	  01~33~32.165 & 30~40~31.82 &	 18.497 $\pm$  0.010 & 19.013 $\pm$  0.014 & 18.762 $\pm$  0.016 & 18.690 $\pm$  0.020 & 18.445 $\pm$  0.022 &   0.14 & 2.322 \\
297...... &	  SM115 &         &         &	  01~33~32.160 & 30~56~05.04 &	 20.142 $\pm$  0.027 & 20.092 $\pm$  0.026 & 19.598 $\pm$  0.026 & 19.340 $\pm$  0.027 & 19.087 $\pm$  0.031 &   0.10 & 3.096 \\
298...... &	  SM116 &         &         &	  01~33~32.345 & 30~38~28.24 &	 22.819 $\pm$  0.252 & 23.218 $\pm$  0.476 & 21.174 $\pm$  0.116 & 20.330 $\pm$  0.085 & 19.483 $\pm$  0.071 &   0.10 & 1.548 \\
299...... &	  SM117 &   PL124 &    SR65 &	  01~33~32.433 & 30~38~24.53 &	 19.530 $\pm$  0.024 & 19.385 $\pm$  0.022 & 18.755 $\pm$  0.017 & 18.437 $\pm$  0.020 & 18.014 $\pm$  0.025 &   0.15 & 2.322 \\
300...... &	  SM118 &   PL125 &    SR66 &	  01~33~32.591 & 30~39~24.53 &	 20.512 $\pm$  0.036 & 20.632 $\pm$  0.042 & 20.077 $\pm$  0.039 & 19.662 $\pm$  0.038 & 19.075 $\pm$  0.029 &   0.09 & 1.290 \\
301...... &	  SM119 &    PL55 &         &	  01~33~32.719 & 30~36~55.27 &	 17.407 $\pm$  0.007 & 18.447 $\pm$  0.016 & 18.232 $\pm$  0.017 & 17.774 $\pm$  0.015 & 18.029 $\pm$  0.028 &   0.10 & 2.580 \\
302...... &	  SM120 &   PL126 &    SR67 &	  01~33~32.765 & 30~31~45.13 &	 20.888 $\pm$  0.070 & 21.103 $\pm$  0.084 & 20.262 $\pm$  0.050 & 19.754 $\pm$  0.045 & 18.969 $\pm$  0.030 &   0.16 & 1.806 \\
303...... &	  SM121 &   PL127 &         &	  01~33~32.968 & 30~49~41.81 &	 19.359 $\pm$  0.019 & 19.160 $\pm$  0.018 & 18.551 $\pm$  0.018 & 18.132 $\pm$  0.016 & 17.645 $\pm$  0.015 &   0.15 & 4.128 \\
304...... &	  SM122 &   PL128 &         &	  01~33~33.273 & 30~48~30.57 &	 19.095 $\pm$  0.016 & 18.980 $\pm$  0.016 & 18.462 $\pm$  0.015 & 18.151 $\pm$  0.016 & 17.678 $\pm$  0.015 &   0.15 & 3.354 \\
305...... &	  SM123 &    PL56 &         &	  01~33~33.572 & 30~36~35.79 &	 19.289 $\pm$  0.033 & 19.494 $\pm$  0.036 & 19.244 $\pm$  0.039 & 19.197 $\pm$  0.053 & 18.868 $\pm$  0.062 &   0.10 & 2.580 \\
306...... &	  SM124 &   PL129 &    SR72 &	  01~33~33.722 & 30~40~02.98 &	 19.459 $\pm$  0.019 & 19.305 $\pm$  0.020 & 18.812 $\pm$  0.022 & 18.512 $\pm$  0.023 & 18.079 $\pm$  0.027 &   0.14 & 1.806 \\
307...... &	  SM126 &    PL57 &         &	  01~33~34.377 & 30~42~01.37 &	 18.674 $\pm$  0.028 & 19.036 $\pm$  0.022 & 18.637 $\pm$  0.019 & 18.300 $\pm$  0.026 & 17.931 $\pm$  0.023 &   0.20 & 2.580 \\
308...... &	  SM127 &   PL130 &         &	  01~33~34.674 & 30~48~21.35 &	 19.549 $\pm$  0.022 & 19.791 $\pm$  0.029 & 19.172 $\pm$  0.025 & 18.821 $\pm$  0.026 & 18.196 $\pm$  0.023 &   0.15 & 3.096 \\
309...... &	  SM130 &   PL131 &         &	  01~33~35.111 & 30~49~00.17 &	 19.988 $\pm$  0.029 & 19.447 $\pm$  0.019 & 18.373 $\pm$  0.013 & 17.692 $\pm$  0.010 & 17.099 $\pm$  0.009 &   0.20 & 2.838 \\
310...... &	  SM131 &   PL132 &         &	  01~33~35.277 & 30~33~11.67 &	 19.579 $\pm$  0.026 & 19.616 $\pm$  0.031 & 19.125 $\pm$  0.029 & 18.728 $\pm$  0.027 & 18.389 $\pm$  0.041 &   0.10 & 2.322 \\
311...... &	  SM132 &   PL133 &    SR74 &	  01~33~35.617 & 30~38~36.77 &	 16.998 $\pm$  0.007 & 17.640 $\pm$  0.009 & 17.511 $\pm$  0.008 & 17.550 $\pm$  0.011 & 17.520 $\pm$  0.013 &   0.10 & 1.548 \\
312...... &	  SM134 &   PL134 &         &	  01~33~36.186 & 30~47~55.22 &	 18.184 $\pm$  0.010 & 18.429 $\pm$  0.012 & 18.067 $\pm$  0.012 & 17.831 $\pm$  0.013 & 17.475 $\pm$  0.014 &   0.20 & 2.064 \\
313...... &	  SM136 &         &    SR78 &	  01~33~36.725 & 30~27~08.19 &	 19.054 $\pm$  0.021 & 19.051 $\pm$  0.020 & 18.560 $\pm$  0.016 & 18.188 $\pm$  0.014 & 17.690 $\pm$  0.011 &   0.12 & 2.580 \\
314...... &	  SM138 &   PL135 &         &	  01~33~36.784 & 30~49~17.69 &	 18.904 $\pm$  0.013 & 19.168 $\pm$  0.016 & 18.651 $\pm$  0.015 & 18.293 $\pm$  0.015 & 17.903 $\pm$  0.014 &   0.20 & 2.838 \\
315...... &	  SM140 &         &         &	  01~33~37.246 & 30~34~14.03 &	 17.118 $\pm$  0.005 & 17.416 $\pm$  0.006 & 17.122 $\pm$  0.006 & 16.936 $\pm$  0.007 & 16.679 $\pm$  0.008 &   0.10 & 2.064 \\
316...... &	  SM141 &         &    SR80 &	  01~33~37.574 & 30~28~04.68 &	 17.788 $\pm$  0.007 & 18.107 $\pm$  0.009 & 17.729 $\pm$  0.008 & 17.484 $\pm$  0.008 & 17.074 $\pm$  0.008 &   0.08 & 2.580 \\
317...... &	  SM142 &    PL58 &         &	  01~33~37.799 & 30~50~32.39 &	 20.312 $\pm$  0.057 & 20.373 $\pm$  0.035 & 19.495 $\pm$  0.026 & 18.878 $\pm$  0.025 & 18.336 $\pm$  0.021 &   0.25 & 2.580 \\
318...... &	  SM143 &         &    SR81 &	  01~33~37.987 & 30~38~02.22 &	 18.405 $\pm$  0.014 & 18.328 $\pm$  0.013 & 17.496 $\pm$  0.009 & 16.971 $\pm$  0.009 & 16.395 $\pm$  0.009 &   0.08 & 2.838 \\
319...... &	  SM144 &   PL136 &         &	  01~33~38.049 & 30~33~05.49 &	 17.449 $\pm$  0.005 & 17.852 $\pm$  0.008 & 17.573 $\pm$  0.008 & 17.448 $\pm$  0.009 & 17.160 $\pm$  0.012 &   0.10 & 2.322 \\
320...... &	  SM145 &   PL137 &         &	  01~33~38.083 & 30~33~17.69 &	 19.261 $\pm$  0.018 & 19.166 $\pm$  0.018 & 18.548 $\pm$  0.015 & 18.242 $\pm$  0.015 & 17.915 $\pm$  0.018 &   0.10 & 2.322 \\
321...... &	  SM146 &    PL59 &         &	  01~33~38.130 & 30~42~22.93 &	 18.760 $\pm$  0.015 & 19.100 $\pm$  0.017 & 18.733 $\pm$  0.020 & 18.461 $\pm$  0.026 & 18.028 $\pm$  0.029 &   0.20 & 3.096 \\
322...... &	  SM149 &         &         &	  01~33~39.435 & 30~55~59.81 &	 19.696 $\pm$  0.019 & 19.679 $\pm$  0.021 & 19.037 $\pm$  0.017 & 18.574 $\pm$  0.017 & 18.095 $\pm$  0.016 &   0.10 & 3.096 \\
323...... &	  SM151 &   PL138 &         &	  01~33~39.486 & 30~48~48.26 &	 19.153 $\pm$  0.012 & 19.101 $\pm$  0.014 & 18.719 $\pm$  0.015 & 18.558 $\pm$  0.018 & 18.301 $\pm$  0.021 &   0.10 & 2.322 \\
324...... &	  SM152 &         &    SR86 &	  01~33~39.699 & 30~31~09.18 &	 16.122 $\pm$  0.004 & 16.652 $\pm$  0.005 & 16.462 $\pm$  0.005 & 16.387 $\pm$  0.006 & 16.186 $\pm$  0.006 &   0.05 & 3.354 \\
325...... &	  SM153 &   PL139 &         &	  01~33~39.712 & 30~32~29.56 &	 20.419 $\pm$  0.078 & 20.537 $\pm$  0.067 & 20.047 $\pm$  0.051 & 19.653 $\pm$  0.044 & 18.814 $\pm$  0.032 &   0.10 & 1.290 \\
326...... &	  SM154 &         &    SR87 &	  01~33~39.933 & 30~38~26.22 &	 15.738 $\pm$  0.004 & 16.316 $\pm$  0.004 & 15.993 $\pm$  0.004 & 15.714 $\pm$  0.004 & 15.209 $\pm$  0.004 &   0.05 & 2.322 \\
327...... &	  SM155 &   PL140 &         &	  01~33~40.088 & 30~21~37.13 &	 20.530 $\pm$  0.072 & 20.421 $\pm$  0.060 & 20.033 $\pm$  0.051 & 19.718 $\pm$  0.043 & 19.386 $\pm$  0.042 &   0.10 & 1.806 \\
328...... &	  SM156 &    PL60 &         &	  01~33~40.370 & 30~43~58.04 &	 16.743 $\pm$  0.004 & 17.268 $\pm$  0.005 & 17.036 $\pm$  0.006 & 16.922 $\pm$  0.007 & 16.649 $\pm$  0.008 &   0.20 & 3.096 \\
329...... &	  SM159 &         &    SR89 &	  01~33~41.190 & 30~29~53.93 &	 19.762 $\pm$  0.028 & 19.855 $\pm$  0.034 & 19.244 $\pm$  0.027 & 18.852 $\pm$  0.027 & 18.393 $\pm$  0.022 &   0.12 & 2.322 \\
330...... &	  SM161 &    PL61 &         &	  01~33~41.535 & 30~42~44.93 &	 18.969 $\pm$  0.016 & 19.144 $\pm$  0.019 & 18.757 $\pm$  0.018 & 18.484 $\pm$  0.019 & 17.985 $\pm$  0.020 &   0.20 & 2.322 \\
331...... &	  SM162 &         &    SR91 &	  01~33~41.550 & 30~30~24.23 &	 18.892 $\pm$  0.017 & 19.445 $\pm$  0.029 & 19.216 $\pm$  0.033 & 18.820 $\pm$  0.034 & 18.186 $\pm$  0.028 &   0.05 & 3.096 \\
332...... &	  SM163 &         &    SR92 &	  01~33~41.594 & 30~28~09.38 &	 19.314 $\pm$  0.022 & 19.889 $\pm$  0.033 & 19.891 $\pm$  0.043 & 19.900 $\pm$  0.058 & 20.002 $\pm$  0.083 &   0.05 & 2.064 \\
333...... &	  SM165 &   PL141 &         &	  01~33~41.591 & 30~48~08.65 &	 19.411 $\pm$  0.025 & 19.411 $\pm$  0.023 & 19.028 $\pm$  0.027 & 18.702 $\pm$  0.028 & 18.354 $\pm$  0.030 &   0.10 & 3.354 \\
334...... &	  SM166 &   PL142 &         &	  01~33~41.908 & 30~49~20.18 &	 21.438 $\pm$  0.078 & 20.996 $\pm$  0.051 & 19.970 $\pm$  0.036 & 19.237 $\pm$  0.026 & 18.399 $\pm$  0.018 &   0.20 & 2.580 \\
335...... &	  SM168 &   PL143 &         &	  01~33~42.706 & 30~43~49.65 &	 18.220 $\pm$  0.014 & 18.666 $\pm$  0.018 & 18.224 $\pm$  0.021 & 17.875 $\pm$  0.022 & 17.461 $\pm$  0.024 &   0.10 & 2.580 \\
336...... &	  SM171 &   PL144 &         &	  01~33~43.742 & 30~40~56.61 &	 17.938 $\pm$  0.011 & 18.784 $\pm$  0.023 & 18.459 $\pm$  0.030 & 17.892 $\pm$  0.024 & 17.849 $\pm$  0.033 &   0.15 & 2.322 \\
337...... &	  SM172 &   PL145 &         &	  01~33~43.864 & 30~32~10.46 &	 17.035 $\pm$  0.005 & 17.520 $\pm$  0.007 & 17.347 $\pm$  0.008 & 17.256 $\pm$  0.009 & 17.107 $\pm$  0.010 &   0.20 & 2.322 \\
338...... &	  SM174 &         &   SR103 &	  01~33~44.055 & 30~30~00.93 &	 18.338 $\pm$  0.010 & 18.633 $\pm$  0.014 & 18.283 $\pm$  0.014 & 18.104 $\pm$  0.013 & 17.717 $\pm$  0.015 &   0.08 & 2.580 \\
339...... &	  SM176 &         &   SR106 &	  01~33~44.509 & 30~37~52.80 &	 17.501 $\pm$  0.007 & 17.762 $\pm$  0.008 & 17.487 $\pm$  0.009 & 17.403 $\pm$  0.013 & 17.183 $\pm$  0.018 &   0.12 & 2.064 \\
340...... &	  SM178 &   PL146 &         &	  01~33~45.043 & 30~47~46.83 &	 17.210 $\pm$  0.005 & 17.081 $\pm$  0.005 & 16.298 $\pm$  0.004 & 15.823 $\pm$  0.004 & 15.311 $\pm$  0.003 &   0.10 & 3.354 \\
341...... &	  SM179 &   PL147 &         &	  01~33~45.130 & 30~49~09.39 &	 19.933 $\pm$  0.022 & 20.018 $\pm$  0.028 & 19.764 $\pm$  0.033 & 19.612 $\pm$  0.041 & 19.323 $\pm$  0.051 &   0.10 & 1.806 \\
342...... &	  SM180 &         &         &	  01~33~45.512 & 30~30~40.71 &	 19.752 $\pm$  0.027 & 19.701 $\pm$  0.028 & 19.060 $\pm$  0.022 & 18.674 $\pm$  0.020 & 17.932 $\pm$  0.014 &   0.10 & 2.580 \\
343...... &	  SM181 &   PL148 &         &	  01~33~45.767 & 30~27~17.47 &	 18.959 $\pm$  0.016 & 18.979 $\pm$  0.017 & 18.573 $\pm$  0.016 & 18.330 $\pm$  0.016 & 17.978 $\pm$  0.015 &   0.10 & 2.580 \\
344...... &	  SM182 &   PL149 &         &	  01~33~46.271 & 30~47~51.21 &	 19.858 $\pm$  0.025 & 19.707 $\pm$  0.023 & 19.022 $\pm$  0.021 & 18.488 $\pm$  0.017 & 17.990 $\pm$  0.015 &   0.10 & 2.838 \\
345...... &	  SM184 &   PL150 &         &	  01~33~46.961 & 30~46~36.32 &	 19.508 $\pm$  0.025 & 19.618 $\pm$  0.028 & 18.759 $\pm$  0.017 & 18.275 $\pm$  0.016 & 17.677 $\pm$  0.012 &   0.20 & 2.580 \\
346...... &	  SM186 &   PL151 &         &	  01~33~48.456 & 30~45~38.72 &	 19.478 $\pm$  0.020 & 19.336 $\pm$  0.020 & 18.625 $\pm$  0.018 & 18.236 $\pm$  0.018 & 17.741 $\pm$  0.015 &   0.10 & 2.580 \\
347...... &	  SM187 &   PL152 &         &	  01~33~48.639 & 30~47~42.62 &	 20.636 $\pm$  0.028 & 20.687 $\pm$  0.031 & 20.261 $\pm$  0.032 & 19.966 $\pm$  0.039 & 19.332 $\pm$  0.035 &   0.10 & 1.548 \\
\end{tabular}
\end{sidewaystable}
\addtocounter{table}{-1}

\begin{sidewaystable}
\centering
\tiny
\caption{(Continued.)}
%\vspace{5mm}
\label{t1.tab}
\begin{tabular}{ccccccccccccc}
\tableline
\tableline
ID & ID & ID & ID & R.A.  & Decl.  & $U$  & $B$ &$V$  & $R_C$   & $I_C$  & $E(B-V)$ & $r_{\rm ap}$ \\
 &  &  &  &  (J2000.0) & (J2000.0) & (mag)  & (mag) & (mag) & (mag) & (mag) & (mag) & (\arcs) \\
\hline
348...... &	  SM189 &   PL153 &         &	  01~33~49.135 & 30~49~01.61 &	 20.587 $\pm$  0.039 & 20.582 $\pm$  0.040 & 19.922 $\pm$  0.035 & 19.511 $\pm$  0.036 & 19.024 $\pm$  0.036 &   0.10 & 2.322 \\
349...... &	  SM190 &   PL154 &         &	  01~33~49.348 & 30~47~12.69 &	 18.197 $\pm$  0.008 & 18.424 $\pm$  0.011 & 18.132 $\pm$  0.011 & 17.965 $\pm$  0.013 & 17.650 $\pm$  0.016 &   0.20 & 2.580 \\
350...... &	  SM191 &   PL155 &   SR122 &	  01~33~49.621 & 30~34~25.88 &	 18.917 $\pm$  0.014 & 18.800 $\pm$  0.014 & 18.316 $\pm$  0.015 & 18.037 $\pm$  0.015 & 17.654 $\pm$  0.015 &   0.12 & 2.322 \\
351...... &	  SM192 &   PL156 &   SR124 &	  01~33~50.195 & 30~34~19.00 &	 19.841 $\pm$  0.044 & 19.831 $\pm$  0.036 & 19.131 $\pm$  0.032 & 18.765 $\pm$  0.034 & 18.139 $\pm$  0.033 &   0.12 & 2.838 \\
352...... &	  SM193 &   PL157 &         &	  01~33~50.285 & 30~31~11.05 &	 21.321 $\pm$  0.086 & 21.656 $\pm$  0.106 & 20.775 $\pm$  0.065 & 20.271 $\pm$  0.067 & 19.780 $\pm$  0.071 &   0.10 & 2.064 \\
353...... &	  SM194 &         &         &	  01~33~50.704 & 30~58~50.36 &	 17.295 $\pm$  0.005 & 17.676 $\pm$  0.007 & 17.488 $\pm$  0.008 & 17.374 $\pm$  0.008 & 17.169 $\pm$  0.010 &   0.10 & 3.870 \\
354...... &	  SM195 &   PL158 &         &	  01~33~50.729 & 30~44~55.75 &	 22.958 $\pm$  0.584 & 21.987 $\pm$  0.252 & 20.732 $\pm$  0.108 & 19.842 $\pm$  0.064 & 18.925 $\pm$  0.046 &   0.15 & 1.806 \\
355...... &	  SM196 &   PL159 &         &	  01~33~50.839 & 30~29~00.03 &	 21.590 $\pm$  0.067 & 21.499 $\pm$  0.066 & 20.874 $\pm$  0.059 & 20.530 $\pm$  0.062 & 20.242 $\pm$  0.066 &   0.10 & 1.806 \\
356...... &	  SM197 &    PL62 &         &	  01~33~50.838 & 30~38~34.64 &	 15.544 $\pm$  0.002 & 16.405 $\pm$  0.004 & 16.384 $\pm$  0.005 & 16.515 $\pm$  0.008 & 16.604 $\pm$  0.014 &   0.15 & 2.064 \\
357...... &	  SM198 &    PL63 &         &	  01~33~50.917 & 30~38~55.54 &	 16.373 $\pm$  0.008 & 17.087 $\pm$  0.010 & 16.779 $\pm$  0.009 & 16.383 $\pm$  0.008 & 15.654 $\pm$  0.006 &   0.15 & 2.064 \\
358...... &	  SM199 &   PL160 &         &	  01~33~50.920 & 30~31~44.88 &	 18.074 $\pm$  0.014 & 18.320 $\pm$  0.017 & 17.891 $\pm$  0.018 & 17.730 $\pm$  0.020 & 17.543 $\pm$  0.027 &   0.10 & 3.354 \\
359...... &	  SM201 &   PL161 &   SR126 &	  01~33~51.239 & 30~34~13.31 &	 18.755 $\pm$  0.015 & 18.877 $\pm$  0.017 & 18.494 $\pm$  0.019 & 18.302 $\pm$  0.022 & 17.913 $\pm$  0.026 &   0.09 & 2.322 \\
360...... &	  SM204 &   PL162 &         &	  01~33~51.778 & 30~31~47.35 &	 18.993 $\pm$  0.019 & 19.402 $\pm$  0.025 & 19.107 $\pm$  0.029 & 19.015 $\pm$  0.038 & 18.903 $\pm$  0.056 &   0.20 & 2.580 \\
361...... &	  SM206 &   PL163 &         &	  01~33~52.137 & 30~29~03.83 &	 18.297 $\pm$  0.010 & 18.126 $\pm$  0.009 & 17.290 $\pm$  0.007 & 16.773 $\pm$  0.006 & 16.277 $\pm$  0.006 &   0.10 & 3.354 \\
362...... &	  SM207 &   PL164 &   SR128 &	  01~33~52.382 & 30~35~00.88 &	 19.445 $\pm$  0.018 & 19.664 $\pm$  0.022 & 19.307 $\pm$  0.023 & 19.081 $\pm$  0.029 & 18.776 $\pm$  0.038 &   0.12 & 1.806 \\
363...... &	  SM208 &   PL165 &   SR129 &	  01~33~52.395 & 30~34~21.27 &	 19.711 $\pm$  0.025 & 19.482 $\pm$  0.022 & 18.882 $\pm$  0.024 & 18.538 $\pm$  0.026 & 18.019 $\pm$  0.024 &   0.12 & 2.838 \\
364...... &	  SM210 &    PL64 &         &	  01~33~52.650 & 30~48~10.26 &	 20.357 $\pm$  0.025 & 20.633 $\pm$  0.032 & 20.288 $\pm$  0.034 & 20.051 $\pm$  0.036 & 19.682 $\pm$  0.039 &   0.10 & 1.548 \\
365...... &	  SM212 &   PL166 &   SR130 &	  01~33~53.424 & 30~33~02.98 &	 19.574 $\pm$  0.026 & 19.776 $\pm$  0.030 & 19.267 $\pm$  0.030 & 18.756 $\pm$  0.029 & 18.221 $\pm$  0.025 &   0.14 & 2.580 \\
366...... &	  SM214 &    PL65 &         &	  01~33~53.664 & 30~48~21.60 &	 17.556 $\pm$  0.006 & 17.889 $\pm$  0.008 & 17.549 $\pm$  0.008 & 17.337 $\pm$  0.009 & 16.997 $\pm$  0.009 &   0.10 & 3.096 \\
367...... &	  SM215 &   PL167 &         &	  01~33~54.113 & 30~33~09.88 &	 16.544 $\pm$  0.009 & 17.464 $\pm$  0.010 & 17.259 $\pm$  0.009 & 17.089 $\pm$  0.013 & 17.268 $\pm$  0.011 &   0.10 & 1.548 \\
368...... &	  SM216 &         &         &	  01~33~54.381 & 30~21~52.00 &	      ......         & 19.095 $\pm$  0.018 & 18.577 $\pm$  0.018 & 18.236 $\pm$  0.018 & 15.343 $\pm$  0.002 &   0.10 & 3.612 \\
369...... &	  SM217 &   PL168 &   SR133 &	  01~33~54.629 & 30~34~48.42 &	 19.103 $\pm$  0.023 & 19.364 $\pm$  0.032 & 18.886 $\pm$  0.041 & 18.612 $\pm$  0.060 & 18.305 $\pm$  0.065 &   0.12 & 3.354 \\
370...... &	  SM218 &    PL66 &         &	  01~33~54.704 & 30~48~43.80 &	 19.411 $\pm$  0.023 & 19.608 $\pm$  0.031 & 19.235 $\pm$  0.040 & 19.090 $\pm$  0.058 & 18.062 $\pm$  0.039 &   0.10 & 2.064 \\
371...... &	  SM219 &   PL169 &         &	  01~33~54.742 & 30~45~28.47 &	 18.357 $\pm$  0.019 & 18.403 $\pm$  0.018 & 17.988 $\pm$  0.018 & 17.686 $\pm$  0.015 & 17.341 $\pm$  0.018 &   0.10 & 2.580 \\
372...... &	  SM220 &   PL170 &   SR134 &	  01~33~54.785 & 30~32~15.90 &	 17.934 $\pm$  0.015 & 18.163 $\pm$  0.014 & 17.886 $\pm$  0.014 & 17.756 $\pm$  0.014 & 17.578 $\pm$  0.018 &   0.08 & 2.580 \\
373...... &	  SM221 &   PL171 &   SR135 &	  01~33~55.041 & 30~32~14.68 &	 18.871 $\pm$  0.030 & 18.941 $\pm$  0.024 & 18.589 $\pm$  0.026 & 18.442 $\pm$  0.031 & 18.254 $\pm$  0.040 &   0.12 & 2.580 \\
374...... &	  SM222 &    PL67 &         &	  01~33~55.154 & 30~47~58.11 &	 16.776 $\pm$  0.005 & 16.981 $\pm$  0.006 & 16.623 $\pm$  0.006 & 16.415 $\pm$  0.006 & 16.070 $\pm$  0.006 &   0.10 & 3.354 \\
375...... &	  SM228 &    PL68 &         &	  01~33~56.168 & 30~38~39.86 &	 17.928 $\pm$  0.011 & 18.473 $\pm$  0.016 & 18.173 $\pm$  0.017 & 17.898 $\pm$  0.019 & 17.336 $\pm$  0.018 &   0.20 & 2.064 \\
376...... &	  SM229 &   PL172 &         &	  01~33~56.200 & 30~45~51.79 &	 19.210 $\pm$  0.030 & 19.419 $\pm$  0.032 & 19.053 $\pm$  0.034 & 18.977 $\pm$  0.047 & 18.606 $\pm$  0.051 &   0.10 & 2.838 \\
377...... &	  SM231 &    PL69 &   SR137 &	  01~33~56.452 & 30~36~10.74 &	 19.427 $\pm$  0.015 & 19.490 $\pm$  0.020 & 18.971 $\pm$  0.019 & 18.670 $\pm$  0.021 & 18.237 $\pm$  0.024 &   0.12 & 1.548 \\
378...... &	  SM232 &   PL173 &         &	  01~33~56.921 & 30~41~38.46 &	 19.908 $\pm$  0.037 & 19.756 $\pm$  0.034 & 19.153 $\pm$  0.035 & 18.636 $\pm$  0.039 & 18.011 $\pm$  0.038 &   0.10 & 2.838 \\
379...... &	  SM233 &    PL70 &         &	  01~33~56.914 & 30~49~26.73 &	 18.963 $\pm$  0.016 & 19.152 $\pm$  0.021 & 18.579 $\pm$  0.019 & 18.187 $\pm$  0.020 & 17.647 $\pm$  0.016 &   0.10 & 3.096 \\
380...... &	  SM234 &         &         &	  01~33~57.129 & 30~50~31.65 &	 19.550 $\pm$  0.026 & 19.532 $\pm$  0.029 & 18.666 $\pm$  0.022 & 18.209 $\pm$  0.019 & 17.612 $\pm$  0.017 &   0.10 & 3.096 \\
381...... &	  SM235 &    PL71 &         &	  01~33~57.075 & 30~48~03.62 &	 19.958 $\pm$  0.035 & 19.738 $\pm$  0.028 & 19.079 $\pm$  0.022 & 18.566 $\pm$  0.022 & 17.970 $\pm$  0.018 &   0.10 & 2.838 \\
382...... &	  SM236 &    PL72 &         &	  01~33~57.136 & 30~40~20.68 &	 19.667 $\pm$  0.037 & 19.696 $\pm$  0.038 & 19.245 $\pm$  0.039 & 18.962 $\pm$  0.042 & 18.699 $\pm$  0.059 &   0.10 & 2.580 \\
383...... &	  SM237 &    PL73 &         &	  01~33~57.265 & 30~39~15.36 &	 18.440 $\pm$  0.012 & 18.583 $\pm$  0.014 & 18.223 $\pm$  0.019 & 17.908 $\pm$  0.026 & 17.584 $\pm$  0.036 &   0.10 & 2.064 \\
384...... &	  SM238 &   PL174 &         &	  01~33~57.341 & 30~41~28.45 &	 19.724 $\pm$  0.016 & 19.634 $\pm$  0.017 & 19.097 $\pm$  0.017 & 18.793 $\pm$  0.021 & 18.350 $\pm$  0.023 &   0.10 & 1.806 \\
385...... &	  SM239 &         &         &	  01~33~57.357 & 30~52~17.95 &	 18.122 $\pm$  0.008 & 18.208 $\pm$  0.011 & 17.904 $\pm$  0.012 & 17.773 $\pm$  0.015 & 17.585 $\pm$  0.018 &   0.10 & 3.354 \\
386...... &	  SM240 &   PL175 &         &	  01~33~57.652 & 30~41~32.58 &	 19.751 $\pm$  0.019 & 19.794 $\pm$  0.022 & 19.356 $\pm$  0.023 & 18.959 $\pm$  0.022 & 18.337 $\pm$  0.019 &   0.10 & 1.806 \\
387...... &	  SM241 &    PL76 &   SR140 &	  01~33~57.830 & 30~35~31.89 &	 18.916 $\pm$  0.037 & 18.945 $\pm$  0.029 & 18.133 $\pm$  0.020 & 17.699 $\pm$  0.021 & 17.172 $\pm$  0.019 &   0.11 & 2.838 \\
388...... &	  SM242 &    PL74 &         &	  01~33~57.829 & 30~49~04.82 &	 18.059 $\pm$  0.009 & 19.351 $\pm$  0.020 & 19.185 $\pm$  0.023 & 18.429 $\pm$  0.019 & 19.082 $\pm$  0.052 &   0.10 & 2.322 \\
389...... &	  SM243 &   PL176 &   SR139 &	  01~33~57.871 & 30~33~25.86 &	 16.525 $\pm$  0.004 & 17.202 $\pm$  0.006 & 17.078 $\pm$  0.006 & 16.971 $\pm$  0.007 & 16.718 $\pm$  0.008 &   0.12 & 2.580 \\
390...... &	  SM245 &   PL177 &         &	  01~33~58.010 & 30~45~45.26 &	 17.237 $\pm$  0.007 & 17.497 $\pm$  0.009 & 17.173 $\pm$  0.009 & 16.945 $\pm$  0.010 & 16.640 $\pm$  0.011 &   0.15 & 3.096 \\
391...... &	  SM246 &    PL77 &         &	  01~33~58.012 & 30~39~26.25 &	 16.925 $\pm$  0.006 & 17.607 $\pm$  0.010 & 17.459 $\pm$  0.013 & 17.392 $\pm$  0.019 & 17.267 $\pm$  0.030 &   0.10 & 3.096 \\
392...... &	  SM247 &    PL78 &         &	  01~33~58.050 & 30~38~15.57 &	 17.698 $\pm$  0.007 & 18.275 $\pm$  0.010 & 18.121 $\pm$  0.016 & 18.037 $\pm$  0.022 & 17.723 $\pm$  0.026 &   0.20 & 2.322 \\
393...... &	  SM248 &    PL79 &         &	  01~33~58.387 & 30~39~15.05 &	 18.429 $\pm$  0.019 & 18.617 $\pm$  0.020 & 18.223 $\pm$  0.027 & 18.034 $\pm$  0.041 & 17.428 $\pm$  0.039 &   0.15 & 3.096 \\
394...... &	  SM249 &    PL80 &         &	  01~33~58.557 & 30~48~42.74 &	 17.968 $\pm$  0.014 & 18.976 $\pm$  0.018 & 18.887 $\pm$  0.021 & 18.375 $\pm$  0.023 & 18.613 $\pm$  0.038 &   0.20 & 2.064 \\
395...... &	  SM250 &   PL178 &         &	  01~33~58.859 & 30~34~43.34 &	 19.204 $\pm$  0.038 & 19.580 $\pm$  0.036 & 19.349 $\pm$  0.038 & 19.368 $\pm$  0.051 & 19.556 $\pm$  0.094 &   0.10 & 1.548 \\
396...... &	  SM251 &    PL81 &         &	  01~33~58.875 & 30~49~11.12 &	 17.537 $\pm$  0.007 & 18.516 $\pm$  0.013 & 18.273 $\pm$  0.014 & 17.704 $\pm$  0.012 & 17.839 $\pm$  0.019 &   0.15 & 2.580 \\
397...... &	  SM252 &    PL82 &         &	  01~33~59.054 & 30~50~05.97 &	 19.250 $\pm$  0.018 & 19.257 $\pm$  0.026 & 18.744 $\pm$  0.023 & 18.437 $\pm$  0.024 & 18.055 $\pm$  0.025 &   0.15 & 2.322 \\
398...... &	  SM253 &    PL83 &         &	  01~33~59.434 & 30~48~26.83 &	 20.722 $\pm$  0.045 & 21.110 $\pm$  0.066 & 20.795 $\pm$  0.071 & 20.679 $\pm$  0.088 & 20.350 $\pm$  0.109 &   0.15 & 1.806 \\
399...... &	  SM254 &   PL179 &         &	  01~33~59.494 & 30~47~29.69 &	 20.290 $\pm$  0.026 & 20.447 $\pm$  0.032 & 19.981 $\pm$  0.028 & 19.557 $\pm$  0.025 & 19.143 $\pm$  0.029 &   0.20 & 1.806 \\
400...... &	  SM255 &   PL180 &         &	  01~33~59.523 & 30~45~49.92 &	 16.311 $\pm$  0.005 & 16.863 $\pm$  0.007 & 16.709 $\pm$  0.008 & 16.618 $\pm$  0.010 & 16.299 $\pm$  0.011 &   0.15 & 3.612 \\
401...... &	  SM256 &   PL181 &         &	  01~33~59.640 & 30~47~38.36 &	 20.408 $\pm$  0.026 & 20.660 $\pm$  0.035 & 20.178 $\pm$  0.033 & 20.017 $\pm$  0.040 & 19.718 $\pm$  0.042 &   0.15 & 1.806 \\
402...... &	  SM257 &   PL182 &         &	  01~33~59.734 & 30~41~24.35 &	 16.763 $\pm$  0.003 & 17.551 $\pm$  0.007 & 17.362 $\pm$  0.007 & 17.338 $\pm$  0.010 & 17.379 $\pm$  0.016 &   0.10 & 2.580 \\
403...... &	  SM258 &    PL84 &         &	  01~33~59.815 & 30~39~45.41 &	 19.355 $\pm$  0.070 & 19.807 $\pm$  0.087 & 19.081 $\pm$  0.073 & 18.564 $\pm$  0.066 & 17.961 $\pm$  0.064 &   0.15 & 2.580 \\
\end{tabular}
\end{sidewaystable}
\addtocounter{table}{-1}

\begin{sidewaystable}
\centering
\tiny
\caption{(Continued.)}
%\vspace{5mm}
\label{t1.tab}
\begin{tabular}{ccccccccccccc}
\tableline
\tableline
ID & ID & ID & ID & R.A.  & Decl.  & $U$  & $B$ &$V$  & $R_C$   & $I_C$  & $E(B-V)$ & $r_{\rm ap}$ \\
 &  &  &  &  (J2000.0) & (J2000.0) & (mag)  & (mag) & (mag) & (mag) & (mag) & (mag) & (\arcs) \\
\hline
404...... &	  SM260 &   PL183 &   SR146 &	  01~34~00.005 & 30~33~54.44 &	 15.451 $\pm$  0.003 & 16.230 $\pm$  0.004 & 16.109 $\pm$  0.004 & 15.920 $\pm$  0.004 & 15.471 $\pm$  0.004 &   0.08 & 3.096 \\
405...... &	  SM261 &    PL85 &         &	  01~34~00.188 & 30~37~47.39 &	 16.553 $\pm$  0.004 & 16.612 $\pm$  0.004 & 15.953 $\pm$  0.003 & 15.650 $\pm$  0.003 & 15.356 $\pm$  0.003 &   0.10 & 2.580 \\
406...... &	  SM262 &    PL86 &         &	  01~34~00.246 & 30~48~36.68 &	 19.569 $\pm$  0.024 & 19.587 $\pm$  0.025 & 19.179 $\pm$  0.025 & 18.818 $\pm$  0.025 & 18.474 $\pm$  0.027 &   0.10 & 2.322 \\
407...... &	  SM264 &   PL184 &         &	  01~34~00.465 & 30~41~23.03 &	 19.214 $\pm$  0.027 & 19.131 $\pm$  0.023 & 18.534 $\pm$  0.021 & 18.040 $\pm$  0.020 & 17.505 $\pm$  0.022 &   0.15 & 2.580 \\
408...... &	  SM265 &   PL185 &         &	  01~34~00.728 & 30~50~09.24 &	 18.925 $\pm$  0.022 & 19.072 $\pm$  0.024 & 18.619 $\pm$  0.025 & 18.228 $\pm$  0.025 & 17.720 $\pm$  0.024 &   0.15 & 3.096 \\
409...... &	  SM266 &   PL186 &         &	  01~34~01.032 & 30~46~58.74 &	 20.705 $\pm$  0.056 & 20.629 $\pm$  0.052 & 19.959 $\pm$  0.042 & 19.681 $\pm$  0.049 & 19.240 $\pm$  0.049 &   0.15 & 2.322 \\
410...... &	  SM267 &   PL187 &         &	  01~34~01.286 & 30~39~23.50 &	 18.397 $\pm$  0.018 & 18.638 $\pm$  0.021 & 18.241 $\pm$  0.022 & 18.136 $\pm$  0.037 & 17.823 $\pm$  0.047 &   0.15 & 2.580 \\
411...... &	  SM270 &    PL87 &         &	  01~34~01.658 & 30~49~44.09 &	 18.347 $\pm$  0.014 & 18.412 $\pm$  0.014 & 17.930 $\pm$  0.013 & 17.635 $\pm$  0.014 & 17.276 $\pm$  0.014 &   0.10 & 3.354 \\
412...... &	  SM271 &   PL188 &   SR147 &	  01~34~01.750 & 30~32~25.81 &	 18.976 $\pm$  0.018 & 19.031 $\pm$  0.021 & 18.528 $\pm$  0.020 & 18.159 $\pm$  0.020 & 17.686 $\pm$  0.020 &   0.09 & 3.354 \\
413...... &	  SM272 &    PL88 &         &	  01~34~01.965 & 30~38~11.13 &	 18.574 $\pm$  0.025 & 18.819 $\pm$  0.027 & 18.512 $\pm$  0.030 & 18.417 $\pm$  0.041 & 18.083 $\pm$  0.053 &   0.20 & 3.354 \\
414...... &	  SM273 &   PL189 &         &	  01~34~01.982 & 30~39~37.92 &	 17.346 $\pm$  0.007 & 17.198 $\pm$  0.006 & 16.399 $\pm$  0.005 & 15.922 $\pm$  0.005 & 15.414 $\pm$  0.006 &   0.15 & 3.354 \\
415...... &	  SM274 &   PL190 &         &	  01~34~02.312 & 30~50~27.81 &	 18.202 $\pm$  0.009 & 19.119 $\pm$  0.018 & 18.956 $\pm$  0.020 & 18.525 $\pm$  0.019 & 18.377 $\pm$  0.024 &   0.15 & 2.838 \\
416...... &	  SM275 &   PL191 &         &	  01~34~02.464 & 30~40~40.68 &	 17.815 $\pm$  0.008 & 17.521 $\pm$  0.008 & 16.545 $\pm$  0.006 & 15.930 $\pm$  0.005 & 15.313 $\pm$  0.005 &   0.10 & 3.354 \\
417...... &	  SM279 &    PL89 &         &	  01~34~02.747 & 30~48~36.57 &	 20.562 $\pm$  0.048 & 19.825 $\pm$  0.027 & 18.011 $\pm$  0.009 & 16.889 $\pm$  0.005 & 15.809 $\pm$  0.003 &   0.15 & 2.580 \\
418...... &	  SM281 &         &         &	  01~34~02.891 & 30~43~20.80 &	 17.163 $\pm$  0.005 & 17.126 $\pm$  0.005 & 16.391 $\pm$  0.004 & 15.936 $\pm$  0.004 & 15.476 $\pm$  0.004 &   0.10 & 3.096 \\
419...... &	  SM282 &   PL192 &         &	  01~34~03.092 & 30~45~35.56 &	 18.759 $\pm$  0.018 & 18.962 $\pm$  0.021 & 18.454 $\pm$  0.023 & 17.997 $\pm$  0.025 & 17.445 $\pm$  0.027 &   0.15 & 3.870 \\
420...... &	  SM284 &         &         &	  01~34~03.102 & 30~52~13.96 &	 17.037 $\pm$  0.005 & 17.171 $\pm$  0.006 & 16.819 $\pm$  0.006 & 16.655 $\pm$  0.006 & 16.392 $\pm$  0.006 &   0.10 & 3.612 \\
421...... &	  SM285 &    PL90 &         &	  01~34~03.098 & 30~48~11.15 &	 19.554 $\pm$  0.040 & 19.786 $\pm$  0.045 & 19.472 $\pm$  0.050 & 19.375 $\pm$  0.068 & 18.515 $\pm$  0.048 &   0.20 & 2.580 \\
422...... &	  SM287 &         &         &	  01~34~03.311 & 30~48~26.73 &	 18.787 $\pm$  0.014 & 19.538 $\pm$  0.027 & 19.319 $\pm$  0.029 & 19.312 $\pm$  0.041 & 19.047 $\pm$  0.057 &   0.10 & 2.580 \\
423...... &	  SM289 &   PL193 &         &	  01~34~03.881 & 30~47~29.19 &	 17.954 $\pm$  0.009 & 18.039 $\pm$  0.011 & 17.556 $\pm$  0.012 & 17.248 $\pm$  0.012 & 16.791 $\pm$  0.012 &   0.10 & 3.612 \\
424...... &	  SM290 &    PL91 &         &	  01~34~04.301 & 30~39~22.87 &	 17.579 $\pm$  0.008 & 18.305 $\pm$  0.014 & 18.088 $\pm$  0.016 & 17.897 $\pm$  0.020 & 17.500 $\pm$  0.021 &   0.10 & 2.322 \\
425...... &	  SM291 &    PL92 &         &	  01~34~04.457 & 30~36~56.22 &	 17.591 $\pm$  0.010 & 18.053 $\pm$  0.012 & 17.755 $\pm$  0.014 & 17.463 $\pm$  0.015 & 16.911 $\pm$  0.014 &   0.10 & 3.612 \\
426...... &	  SM292 &   PL194 &         &	  01~34~04.737 & 30~49~18.08 &	 18.654 $\pm$  0.013 & 19.059 $\pm$  0.017 & 18.745 $\pm$  0.018 & 18.423 $\pm$  0.017 & 17.979 $\pm$  0.016 &   0.15 & 2.580 \\
427...... &	  SM294 &   PL196 &         &	  01~34~05.065 & 30~49~42.70 &	 20.679 $\pm$  0.073 & 20.658 $\pm$  0.067 & 19.685 $\pm$  0.040 & 19.041 $\pm$  0.029 & 18.448 $\pm$  0.027 &   0.15 & 2.580 \\
428...... &	  SM296 &         &         &	  01~34~05.223 & 30~57~01.30 &	 19.458 $\pm$  0.024 & 19.485 $\pm$  0.033 & 18.797 $\pm$  0.022 & 18.372 $\pm$  0.019 & 17.773 $\pm$  0.015 &   0.10 & 4.128 \\
429...... &	  SM298 &   PL198 &         &	  01~34~05.824 & 30~49~56.88 &	 17.582 $\pm$  0.006 & 17.952 $\pm$  0.008 & 17.728 $\pm$  0.008 & 17.646 $\pm$  0.009 & 17.412 $\pm$  0.009 &   0.15 & 2.580 \\
430...... &	  SM300 &    PL93 &         &	  01~34~06.302 & 30~37~26.16 &	 18.696 $\pm$  0.024 & 18.738 $\pm$  0.023 & 17.960 $\pm$  0.017 & 17.532 $\pm$  0.019 & 17.090 $\pm$  0.022 &   0.10 & 3.096 \\
431...... &	  SM301 &    PL94 &         &	  01~34~06.387 & 30~37~30.45 &	 18.934 $\pm$  0.032 & 18.797 $\pm$  0.030 & 18.238 $\pm$  0.029 & 17.678 $\pm$  0.024 & 17.025 $\pm$  0.019 &   0.10 & 3.354 \\
432...... &	  SM302 &   PL199 &         &	  01~34~06.566 & 30~50~18.28 &	 18.823 $\pm$  0.011 & 19.281 $\pm$  0.019 & 19.021 $\pm$  0.022 & 18.881 $\pm$  0.027 & 18.712 $\pm$  0.039 &   0.10 & 2.322 \\
433...... &	  SM303 &    PL95 &         &	  01~34~06.654 & 30~48~56.13 &	 16.506 $\pm$  0.005 & 17.563 $\pm$  0.009 & 17.224 $\pm$  0.011 & 16.807 $\pm$  0.011 & 17.161 $\pm$  0.020 &   0.10 & 4.902 \\
434...... &	  SM304 &   PL200 &         &	  01~34~06.759 & 30~48~32.81 &	 18.884 $\pm$  0.017 & 19.121 $\pm$  0.021 & 18.796 $\pm$  0.022 & 18.638 $\pm$  0.027 & 18.327 $\pm$  0.034 &   0.20 & 3.096 \\
435...... &	  SM305 &   PL201 &         &	  01~34~06.780 & 30~47~27.12 &	 16.199 $\pm$  0.016 & 17.079 $\pm$  0.014 & 17.049 $\pm$  0.017 & 17.149 $\pm$  0.029 & 17.129 $\pm$  0.015 &   0.20 & 1.548 \\
436...... &	  SM306 &         &   SR154 &	  01~34~06.973 & 30~32~00.22 &	 18.946 $\pm$  0.013 & 18.845 $\pm$  0.013 & 18.317 $\pm$  0.014 & 17.959 $\pm$  0.014 & 17.502 $\pm$  0.014 &   0.08 & 3.612 \\
437...... &	  SM307 &   PL202 &         &	  01~34~07.008 & 30~50~57.49 &	 19.410 $\pm$  0.021 & 19.467 $\pm$  0.025 & 19.161 $\pm$  0.030 & 19.146 $\pm$  0.040 & 18.737 $\pm$  0.042 &   0.15 & 2.838 \\
438...... &	  SM308 &   PL203 &         &	  01~34~07.006 & 30~49~24.43 &	 17.669 $\pm$  0.007 & 18.068 $\pm$  0.009 & 17.829 $\pm$  0.010 & 17.688 $\pm$  0.012 & 17.292 $\pm$  0.012 &   0.20 & 3.612 \\
439...... &	  SM310 &    PL96 &         &	  01~34~07.253 & 30~38~29.58 &	 19.074 $\pm$  0.021 & 19.304 $\pm$  0.023 & 18.790 $\pm$  0.027 & 18.367 $\pm$  0.029 & 17.887 $\pm$  0.034 &   0.10 & 2.580 \\
440...... &	  SM311 &   PL204 &         &	  01~34~07.348 & 30~47~41.68 &	 20.539 $\pm$  0.084 & 20.472 $\pm$  0.073 & 19.617 $\pm$  0.050 & 19.150 $\pm$  0.045 & 18.660 $\pm$  0.036 &   0.20 & 2.580 \\
441...... &	  SM312 &   PL205 &         &	  01~34~07.489 & 30~50~11.05 &	 18.566 $\pm$  0.013 & 18.957 $\pm$  0.020 & 18.811 $\pm$  0.026 & 18.599 $\pm$  0.029 & 18.278 $\pm$  0.035 &   0.10 & 3.096 \\
442...... &	  SM316 &   PL206 &         &	  01~34~08.018 & 30~38~38.29 &	 17.512 $\pm$  0.007 & 17.286 $\pm$  0.006 & 16.376 $\pm$  0.004 & 15.839 $\pm$  0.004 & 15.282 $\pm$  0.004 &   0.10 & 3.354 \\
443...... &	  SM320 &    PL97 &         &	  01~34~08.500 & 30~39~02.40 &	 15.853 $\pm$  0.002 & 16.482 $\pm$  0.004 & 16.342 $\pm$  0.004 & 16.270 $\pm$  0.005 & 16.022 $\pm$  0.006 &   0.10 & 2.580 \\
444...... &	  SM321 &    PL98 &         &	  01~34~08.609 & 30~39~22.80 &	 17.585 $\pm$  0.010 & 17.803 $\pm$  0.011 & 17.373 $\pm$  0.012 & 17.125 $\pm$  0.016 & 16.816 $\pm$  0.020 &   0.10 & 3.096 \\
445...... &	  SM322 &    PL99 &         &	  01~34~08.733 & 30~42~55.24 &	 19.355 $\pm$  0.022 & 19.232 $\pm$  0.021 & 18.813 $\pm$  0.027 & 18.523 $\pm$  0.033 & 18.252 $\pm$  0.050 &   0.10 & 3.354 \\
446...... &	  SM323 &   PL207 &         &	  01~34~08.762 & 30~48~16.27 &	 19.271 $\pm$  0.027 & 19.323 $\pm$  0.030 & 18.972 $\pm$  0.037 & 18.711 $\pm$  0.041 & 18.219 $\pm$  0.045 &   0.15 & 3.354 \\
447...... &	  SM327 &   PL208 &         &	  01~34~09.702 & 30~21~29.96 &	 18.766 $\pm$  0.010 & 18.890 $\pm$  0.010 & 18.618 $\pm$  0.010 & 18.451 $\pm$  0.011 & 18.116 $\pm$  0.011 &   0.05 & 2.838 \\
448...... &	  SM328 &         &         &	  01~34~09.769 & 30~52~06.13 &	 19.210 $\pm$  0.018 & 19.556 $\pm$  0.026 & 19.374 $\pm$  0.029 & 19.137 $\pm$  0.030 & 18.829 $\pm$  0.029 &   0.10 & 2.580 \\
449...... &	  SM329 &         &         &	  01~34~10.099 & 30~45~29.44 &	 17.807 $\pm$  0.010 & 17.896 $\pm$  0.012 & 17.476 $\pm$  0.012 & 17.336 $\pm$  0.016 & 16.943 $\pm$  0.016 &   0.10 & 4.386 \\
450...... &	  SM333 &         &         &	  01~34~10.935 & 30~40~30.03 &	 18.339 $\pm$  0.012 & 18.259 $\pm$  0.013 & 17.730 $\pm$  0.013 & 17.377 $\pm$  0.014 & 16.896 $\pm$  0.016 &   0.10 & 3.354 \\
451...... &	  SM335 &         &         &	  01~34~11.355 & 30~41~27.95 &	 18.379 $\pm$  0.015 & 18.369 $\pm$  0.015 & 17.880 $\pm$  0.015 & 17.640 $\pm$  0.018 & 17.295 $\pm$  0.021 &   0.10 & 4.128 \\
452...... &	  SM338 &   PL100 &         &	  01~34~11.795 & 30~42~20.05 &	 19.987 $\pm$  0.019 & 20.087 $\pm$  0.024 & 19.689 $\pm$  0.022 & 19.453 $\pm$  0.025 & 19.295 $\pm$  0.039 &   0.25 & 1.290 \\
453...... &	  SM342 &   PL101 &         &	  01~34~13.687 & 30~43~18.44 &	 18.954 $\pm$  0.013 & 19.327 $\pm$  0.018 & 19.058 $\pm$  0.021 & 19.088 $\pm$  0.031 & 19.181 $\pm$  0.054 &   0.15 & 2.580 \\
454...... &	  SM345 &         &         &	  01~34~13.841 & 30~19~47.31 &	 18.767 $\pm$  0.008 & 18.754 $\pm$  0.009 & 18.385 $\pm$  0.009 & 18.126 $\pm$  0.009 & 17.555 $\pm$  0.008 &   0.10 & 3.354 \\
455...... &	  SM346 &         &   SR159 &	  01~34~13.979 & 30~27~58.99 &	 18.443 $\pm$  0.011 & 18.870 $\pm$  0.014 & 18.135 $\pm$  0.011 & 17.386 $\pm$  0.008 & 16.427 $\pm$  0.005 &   0.05 & 3.870 \\
456...... &	  SM347 &   PL209 &         &	  01~34~13.996 & 30~39~29.47 &	 18.829 $\pm$  0.015 & 18.795 $\pm$  0.016 & 18.367 $\pm$  0.016 & 18.169 $\pm$  0.019 & 17.751 $\pm$  0.021 &   0.15 & 2.838 \\
457...... &	  SM350 &   PL210 &         &	  01~34~14.236 & 30~39~58.48 &	 18.546 $\pm$  0.013 & 18.477 $\pm$  0.015 & 17.911 $\pm$  0.015 & 17.666 $\pm$  0.016 & 17.245 $\pm$  0.018 &   0.10 & 3.354 \\
458...... &	  SM351 &   PL102 &         &	  01~34~14.647 & 30~32~35.19 &	 18.653 $\pm$  0.010 & 18.589 $\pm$  0.010 & 18.153 $\pm$  0.011 & 17.912 $\pm$  0.010 & 17.582 $\pm$  0.012 &   0.20 & 3.096 \\
459...... &	  SM353 &   PL103 &         &	  01~34~15.037 & 30~41~19.14 &	 17.638 $\pm$  0.007 & 17.835 $\pm$  0.009 & 17.431 $\pm$  0.010 & 17.227 $\pm$  0.012 & 16.921 $\pm$  0.014 &   0.10 & 4.128 \\
\end{tabular}
\end{sidewaystable}
\addtocounter{table}{-1}

\begin{sidewaystable}
\centering
\tiny
\caption{(Continued.)}
%\vspace{5mm}
\label{t1.tab}
\begin{tabular}{ccccccccccccc}
\tableline
\tableline
ID & ID & ID & ID & R.A.  & Decl.  & $U$  & $B$ &$V$  & $R_C$   & $I_C$  & $E(B-V)$ & $r_{\rm ap}$ \\
 &  &  &  &  (J2000.0) & (J2000.0) & (mag)  & (mag) & (mag) & (mag) & (mag) & (mag) & (\arcs) \\
\hline
460...... &	  SM355 &   PL104 &         &	  01~34~15.510 & 30~42~11.35 &	 18.020 $\pm$  0.011 & 18.224 $\pm$  0.013 & 17.785 $\pm$  0.013 & 17.447 $\pm$  0.014 & 16.933 $\pm$  0.012 &   0.10 & 3.612 \\
461...... &	  SM358 &         &         &	  01~34~16.370 & 30~47~43.10 &	 18.840 $\pm$  0.014 & 18.758 $\pm$  0.015 & 18.208 $\pm$  0.015 & 17.919 $\pm$  0.016 & 17.538 $\pm$  0.017 &   0.10 & 3.096 \\
462...... &	  SM359 &   PL211 &         &	  01~34~16.358 & 30~37~49.12 &	 17.678 $\pm$  0.015 & 18.024 $\pm$  0.014 & 17.637 $\pm$  0.012 & 17.433 $\pm$  0.012 & 17.115 $\pm$  0.011 &   0.10 & 2.838 \\
463...... &	  SM360 &   PL212 &         &	  01~34~16.546 & 30~40~28.92 &	 19.858 $\pm$  0.029 & 19.676 $\pm$  0.025 & 19.191 $\pm$  0.026 & 18.880 $\pm$  0.030 & 18.516 $\pm$  0.037 &   0.10 & 2.838 \\
464...... &	  SM361 &   PL213 &         &	  01~34~17.549 & 30~42~36.57 &	 20.099 $\pm$  0.024 & 20.235 $\pm$  0.033 & 19.857 $\pm$  0.036 & 19.474 $\pm$  0.038 & 18.787 $\pm$  0.032 &   0.10 & 2.580 \\
465...... &	  SM367 &   PL214 &         &	  01~34~18.575 & 30~44~47.83 &	 21.749 $\pm$  0.119 & 21.044 $\pm$  0.066 & 19.459 $\pm$  0.021 & 18.620 $\pm$  0.015 & 17.869 $\pm$  0.012 &   0.10 & 1.548 \\
466...... &	  SM371 &   PL215 &         &	  01~34~19.882 & 30~36~12.84 &	 17.037 $\pm$  0.004 & 17.387 $\pm$  0.006 & 17.123 $\pm$  0.006 & 16.978 $\pm$  0.007 & 16.765 $\pm$  0.007 &   0.15 & 3.096 \\
467...... &	  SM372 &   PL216 &         &	  01~34~20.146 & 30~39~33.21 &	 18.839 $\pm$  0.012 & 18.832 $\pm$  0.013 & 18.407 $\pm$  0.016 & 18.183 $\pm$  0.018 & 17.865 $\pm$  0.017 &   0.10 & 2.838 \\
468...... &	  SM375 &   PL217 &         &	  01~34~21.403 & 30~39~40.13 &	 20.193 $\pm$  0.034 & 20.161 $\pm$  0.034 & 19.518 $\pm$  0.035 & 19.151 $\pm$  0.035 & 18.591 $\pm$  0.037 &   0.10 & 2.580 \\
469...... &	  SM376 &   PL218 &         &	  01~34~21.573 & 30~36~45.73 &	 18.803 $\pm$  0.013 & 19.019 $\pm$  0.016 & 18.594 $\pm$  0.018 & 18.395 $\pm$  0.021 & 18.126 $\pm$  0.029 &   0.10 & 2.580 \\
470...... &	  SM377 &   PL219 &         &	  01~34~21.997 & 30~44~39.21 &	 18.783 $\pm$  0.015 & 19.026 $\pm$  0.019 & 18.626 $\pm$  0.019 & 18.553 $\pm$  0.025 & 18.249 $\pm$  0.030 &   0.10 & 2.580 \\
471...... &	  SM382 &   PL220 &         &	  01~34~23.035 & 30~37~39.88 &	 19.351 $\pm$  0.018 & 19.496 $\pm$  0.022 & 19.065 $\pm$  0.024 & 18.738 $\pm$  0.028 & 18.389 $\pm$  0.030 &   0.15 & 3.096 \\
472...... &	  SM383 &   PL221 &         &	  01~34~23.130 & 30~43~46.43 &	 19.956 $\pm$  0.026 & 20.004 $\pm$  0.030 & 19.553 $\pm$  0.030 & 19.088 $\pm$  0.027 & 18.519 $\pm$  0.024 &   0.10 & 2.580 \\
473...... &	  SM384 &         &         &	  01~34~23.517 & 30~25~58.26 &	 18.197 $\pm$  0.007 & 18.226 $\pm$  0.008 & 17.842 $\pm$  0.007 & 17.630 $\pm$  0.007 & 17.346 $\pm$  0.007 &   0.10 & 3.096 \\
474...... &	  SM385 &         &         &	  01~34~24.533 & 30~53~05.51 &	 19.261 $\pm$  0.018 & 19.014 $\pm$  0.017 & 18.199 $\pm$  0.013 & 17.668 $\pm$  0.009 & 17.119 $\pm$  0.008 &   0.10 & 3.354 \\
475...... &	  SM387 &         &         &	  01~34~25.373 & 30~41~28.38 &	 18.324 $\pm$  0.010 & 18.181 $\pm$  0.010 & 17.507 $\pm$  0.008 & 17.095 $\pm$  0.008 & 16.660 $\pm$  0.008 &   0.10 & 3.354 \\
476...... &	  SM388 &   PL222 &         &	  01~34~25.485 & 30~36~56.84 &	 18.438 $\pm$  0.011 & 18.581 $\pm$  0.013 & 18.188 $\pm$  0.014 & 17.895 $\pm$  0.015 & 17.342 $\pm$  0.013 &   0.10 & 3.096 \\
477...... &	  SM389 &   PL223 &         &	  01~34~26.302 & 30~37~23.39 &	 18.182 $\pm$  0.009 & 18.554 $\pm$  0.011 & 18.181 $\pm$  0.011 & 17.819 $\pm$  0.011 & 17.382 $\pm$  0.010 &   0.10 & 3.096 \\
478...... &	  SM392 &   PL224 &         &	  01~34~27.124 & 30~36~42.41 &	 18.136 $\pm$  0.010 & 18.181 $\pm$  0.010 & 17.697 $\pm$  0.010 & 17.423 $\pm$  0.009 & 17.084 $\pm$  0.009 &   0.10 & 3.096 \\
479...... &	  SM393 &         &         &	  01~34~27.608 & 30~55~53.37 &	 20.119 $\pm$  0.030 & 20.130 $\pm$  0.035 & 19.653 $\pm$  0.032 & 19.565 $\pm$  0.038 & 19.197 $\pm$  0.039 &   0.10 & 2.838 \\
480...... &	  SM395 &   PL225 &         &	  01~34~28.177 & 30~36~17.13 &	 15.911 $\pm$  0.003 & 16.211 $\pm$  0.004 & 15.938 $\pm$  0.003 & 15.778 $\pm$  0.003 & 15.540 $\pm$  0.003 &   0.10 & 2.838 \\
481...... &	  SM396 &   PL226 &         &	  01~34~28.475 & 30~37~56.12 &	 19.049 $\pm$  0.013 & 19.210 $\pm$  0.015 & 18.900 $\pm$  0.017 & 18.748 $\pm$  0.021 & 18.469 $\pm$  0.023 &   0.15 & 2.580 \\
482...... &	  SM399 &   PL227 &         &	  01~34~29.043 & 30~38~05.24 &	 19.051 $\pm$  0.016 & 19.428 $\pm$  0.021 & 18.999 $\pm$  0.022 & 18.618 $\pm$  0.021 & 18.251 $\pm$  0.022 &   0.10 & 3.096 \\
483...... &	  SM400 &         &         &	  01~34~29.143 & 30~53~20.59 &	 19.095 $\pm$  0.014 & 18.963 $\pm$  0.015 & 18.382 $\pm$  0.014 & 18.084 $\pm$  0.013 & 17.659 $\pm$  0.012 &   0.10 & 3.096 \\
484...... &	  SM401 &         &         &	  01~34~29.292 & 30~56~06.10 &	 19.206 $\pm$  0.016 & 18.938 $\pm$  0.016 & 18.266 $\pm$  0.013 & 17.818 $\pm$  0.013 & 17.198 $\pm$  0.010 &   0.10 & 3.354 \\
485...... &	  SM402 &   PL228 &         &	  01~34~30.226 & 30~38~12.97 &	 17.848 $\pm$  0.007 & 17.880 $\pm$  0.008 & 17.127 $\pm$  0.007 & 16.618 $\pm$  0.005 & 16.108 $\pm$  0.005 &   0.10 & 3.870 \\
486...... &	  SM405 &   PL229 &         &	  01~34~31.047 & 30~37~41.05 &	 19.554 $\pm$  0.023 & 20.009 $\pm$  0.029 & 19.904 $\pm$  0.034 & 19.964 $\pm$  0.048 & 19.875 $\pm$  0.066 &   0.10 & 2.580 \\
487...... &	  SM406 &   PL230 &         &	  01~34~31.721 & 30~39~14.71 &	 20.434 $\pm$  0.027 & 20.669 $\pm$  0.034 & 20.261 $\pm$  0.036 & 19.850 $\pm$  0.036 & 19.425 $\pm$  0.034 &   0.10 & 1.806 \\
488...... &	  SM409 &   PL231 &         &	  01~34~32.884 & 30~38~11.96 &	 19.803 $\pm$  0.026 & 19.679 $\pm$  0.025 & 19.030 $\pm$  0.023 & 18.489 $\pm$  0.020 & 17.848 $\pm$  0.018 &   0.10 & 3.354 \\
489...... &	  SM410 &   PL232 &         &	  01~34~33.073 & 30~37~36.26 &	 18.106 $\pm$  0.009 & 18.473 $\pm$  0.011 & 18.291 $\pm$  0.013 & 18.116 $\pm$  0.014 & 17.861 $\pm$  0.015 &   0.10 & 3.096 \\
490...... &	  SM411 &   PL233 &         &	  01~34~33.099 & 30~38~14.19 &	 19.738 $\pm$  0.026 & 19.629 $\pm$  0.024 & 18.983 $\pm$  0.020 & 18.637 $\pm$  0.021 & 18.317 $\pm$  0.025 &   0.10 & 3.354 \\
491...... &	  SM412 &   PL234 &         &	  01~34~33.169 & 30~38~26.66 &	 20.160 $\pm$  0.029 & 20.226 $\pm$  0.031 & 19.604 $\pm$  0.031 & 19.095 $\pm$  0.027 & 18.610 $\pm$  0.025 &   0.10 & 2.838 \\
492...... &	  SM413 &   PL235 &         &	  01~34~33.714 & 30~39~15.67 &	 18.364 $\pm$  0.011 & 18.663 $\pm$  0.013 & 18.276 $\pm$  0.013 & 17.986 $\pm$  0.015 & 17.428 $\pm$  0.013 &   0.10 & 3.354 \\
493...... &	  SM416 &   PL236 &         &	  01~34~35.310 & 30~38~29.98 &	 19.976 $\pm$  0.026 & 19.708 $\pm$  0.021 & 18.794 $\pm$  0.015 & 18.274 $\pm$  0.012 & 17.849 $\pm$  0.013 &   0.10 & 3.354 \\
494...... &	  SM419 &   PL237 &         &	  01~34~38.967 & 30~38~51.88 &	 18.272 $\pm$  0.006 & 18.951 $\pm$  0.010 & 18.582 $\pm$  0.009 & 18.340 $\pm$  0.009 & 17.979 $\pm$  0.009 &   0.10 & 2.322 \\
495...... &	  SM420 &   PL238 &         &	  01~34~40.411 & 30~46~01.36 &	 15.336 $\pm$  0.004 & 16.076 $\pm$  0.005 & 15.885 $\pm$  0.004 & 15.626 $\pm$  0.004 & 14.806 $\pm$  0.002 &   0.20 & 3.612 \\
496...... &	  SM421 &         &         &	  01~34~40.660 & 30~49~47.28 &	 17.835 $\pm$  0.008 & 18.077 $\pm$  0.011 & 17.800 $\pm$  0.011 & 17.625 $\pm$  0.012 & 17.433 $\pm$  0.014 &   0.10 & 3.612 \\
497...... &	  SM422 &   PL239 &         &	  01~34~40.721 & 30~53~01.87 &	 19.938 $\pm$  0.026 & 19.932 $\pm$  0.030 & 19.416 $\pm$  0.029 & 18.915 $\pm$  0.023 & 17.946 $\pm$  0.014 &   0.05 & 3.096 \\
498...... &	  SM425 &         &         &	  01~34~42.791 & 30~49~19.18 &	 19.814 $\pm$  0.043 & 19.777 $\pm$  0.035 & 18.950 $\pm$  0.027 & 18.413 $\pm$  0.019 & 17.856 $\pm$  0.018 &   0.10 & 3.354 \\
499...... &	  SM426 &   PL240 &         &	  01~34~43.194 & 30~52~19.20 &	 22.137 $\pm$  0.104 & 21.896 $\pm$  0.102 & 20.735 $\pm$  0.050 & 20.246 $\pm$  0.043 & 19.623 $\pm$  0.033 &   0.15 & 2.064 \\
500...... &	  SM427 &         &         &	  01~34~43.735 & 30~47~38.08 &	 17.777 $\pm$  0.007 & 17.748 $\pm$  0.007 & 17.247 $\pm$  0.006 & 16.968 $\pm$  0.006 & 16.583 $\pm$  0.005 &   0.10 & 3.096 \\
501...... &	  SM428 &   PL241 &         &	  01~34~44.202 & 30~52~18.98 &	 18.565 $\pm$  0.011 & 18.427 $\pm$  0.012 & 17.578 $\pm$  0.009 & 17.073 $\pm$  0.007 & 16.595 $\pm$  0.006 &   0.15 & 3.870 \\
502...... &	  SM429 &         &         &	  01~34~45.095 & 30~50~33.41 &	 20.163 $\pm$  0.026 & 19.917 $\pm$  0.028 & 19.050 $\pm$  0.018 & 18.507 $\pm$  0.016 & 18.055 $\pm$  0.013 &   0.10 & 3.096 \\
503...... &	  SM432 &   PL242 &         &	  01~34~45.911 & 30~53~04.36 &	 20.437 $\pm$  0.031 & 20.518 $\pm$  0.040 & 19.939 $\pm$  0.035 & 19.567 $\pm$  0.029 & 19.383 $\pm$  0.030 &   0.05 & 2.580 \\
504...... &	  SM436 &         &         &	  01~34~46.780 & 30~49~16.04 &	 19.279 $\pm$  0.015 & 19.316 $\pm$  0.019 & 18.911 $\pm$  0.018 & 18.760 $\pm$  0.021 & 18.483 $\pm$  0.022 &   0.10 & 2.838 \\
505...... &	  SM438 &         &         &	  01~34~49.621 & 30~21~55.50 &	 16.820 $\pm$  0.004 & 16.813 $\pm$  0.004 & 16.120 $\pm$  0.004 & 15.707 $\pm$  0.003 & 15.317 $\pm$  0.003 &   0.10 & 3.870 \\
506...... &	  SM439 &         &         &	  01~34~50.138 & 30~47~04.26 &	 16.581 $\pm$  0.003 & 16.811 $\pm$  0.004 & 16.501 $\pm$  0.004 & 16.311 $\pm$  0.004 & 16.037 $\pm$  0.004 &   0.10 & 3.612 \\
507...... &	  SM441 &         &         &	  01~34~52.224 & 30~50~05.53 &	 19.384 $\pm$  0.024 & 19.293 $\pm$  0.025 & 19.001 $\pm$  0.026 & 18.771 $\pm$  0.026 & 18.598 $\pm$  0.028 &   0.10 & 4.128 \\
508...... &	  SM443 &         &         &	  01~34~53.179 & 30~51~47.86 &	 20.297 $\pm$  0.031 & 20.298 $\pm$  0.037 & 19.783 $\pm$  0.033 & 19.562 $\pm$  0.033 & 19.312 $\pm$  0.033 &   0.10 & 3.096 \\
509...... &	  SM446 &         &         &	  01~35~01.556 & 30~51~26.91 &	 19.938 $\pm$  0.027 & 19.939 $\pm$  0.032 & 19.397 $\pm$  0.027 & 19.011 $\pm$  0.024 & 18.500 $\pm$  0.020 &   0.10 & 3.612 \\
510...... &	  SM447 &         &         &	  01~35~04.710 & 30~46~10.68 &	 19.023 $\pm$  0.012 & 19.120 $\pm$  0.015 & 18.605 $\pm$  0.013 & 18.253 $\pm$  0.017 & 17.213 $\pm$  0.031 &   0.10 & 3.354 \\
511...... &	  SM449 &         &         &	  01~35~18.248 & 30~49~53.92 &	 19.160 $\pm$  0.014 & 18.962 $\pm$  0.015 & 18.255 $\pm$  0.014 & 17.777 $\pm$  0.011 & 17.286 $\pm$  0.010 &   0.10 & 5.160 \\
\end{tabular}
\end{sidewaystable}

\end{document}